\documentclass[12pt]{article}

\usepackage{amssymb}
\usepackage{amsmath}
\usepackage{amsbsy} 

\usepackage{graphics}

\begin{document}%

\title{\bf OPE coefficient functions in terms of composite
operators only. Singlet case}

\author{A.V. Kisselev\thanks{E-mail: alexandre.kisselev@ihep.ru} \\
\small Institute for High Energy Physics, 142281 Protvino, Russia}

\date{}

\maketitle

\thispagestyle{empty}

\bigskip

\begin{abstract}
A method for calculating coefficient functions of the operator
product expansion, which was previously derived for the nonsinglet
case, is generalized for the singlet coefficient functions. The
resulting formula defines coefficient functions entirely in terms of
corresponding singlet composite operators without applying to
elementary (quark and gluon) fields.  Both ``diagonal'' and
``nondiagonal'' gluon coefficient functions in the product expansion
of two electromagnetic currents are calculated in QCD. Their
renormalization properties are studied.
\end{abstract}
\clearpage


\section{Introduction}

The light-cone (LC) operator product expansion
(OPE)~\cite{Frishman:70} (see also \cite{Geyer:79, Yndurain})
permanently receives much attention, since it enables us to separate
contributions to cross sections coming from large and small
distances in a variety of processes. It was proposed as a
generalization of the OPE at short distances~\cite{Wilson:69} in
order to describe deep inelastic scattering (DIS) of leptons off
nucleons.

In Refs.~\cite{Anikin:78} the T-product of two scalar currents near
the LC was defined in term of so-called bi-local light-ray composite
fields. Later on it was shown that the local LC expansion can be
obtained by performing a Taylor expansion of the nonlocal
one~\cite{Geyer:85}. The nonlocal expansion is more general, but in
the present paper we restrict ourselves to considering standard
local OPE.

The OPE for the $\mathrm{T}$-product of two electromagnetic currents
is of particular importance to practical application. It can be
written in the form (see, for instance, \cite{Nonsinglet_case}):
\begin{align}\label{OPE}
\mbox{T} J^{\mathrm{em}}_{\mu}(x) J^{\mathrm{em}}_{\nu}(0) & = -
g_{\mu\nu} \bigg\{ \sum\limits_{m=2}^{\infty} \sum\limits_{l=1}^m
C^{NS}_{m,l}(x^2) \, \frac{i^m}{m!} \, x^{\mu_1} \ldots x^{\mu_m}
\nonumber \\
& \times  \frac{1}{6} \left[ O^{3,\,m,l}_{NS,\,\mu_1 \ldots
\mu_m}(0) + \frac{1}{\sqrt{3}} O^{8,\,m,l}_{NS,\,\mu_1 \ldots
\mu_m}(0) \right]
\nonumber \\
& + \frac{2}{9} \bigg[ \sum\limits_{m=2}^{\infty}
\sum\limits_{l=1}^m C^{F}_{m,l}(x^2) \, \frac{i^m}{m!} \, x^{\mu_1}
\ldots x^{\mu_m} O^{m,l}_{F,\,\mu_1 \ldots \mu_m}(0)
\nonumber \\
& + \sum\limits_{m=2}^{\infty} \sum\limits_{l=1}^{m-1}
C^{V}_{m,l}(x^2) \, \frac{i^m}{m!} \, x^{\mu_1} \ldots x^{\mu_m}
O^{m,l}_{V,\,\mu_1 \ldots \mu_m}(0) \bigg] \bigg\} + \cdots \, ,
\end{align}
where the dots denote contributions from other Lorentz structures.
This expansion contains both nonsinglet (triplet
$O^{3,\,m,l}_{NS,\,\mu_1 \ldots \mu_m}$ and octet
$O^{8,\,m,l}_{NS,\,\mu_1 \ldots \mu_m}$) composite operators and
singlet (quark $O^{m,l}_{F,\,\mu_1 \ldots \mu_m}$ and gluon
$O^{m,l}_{V,\,\mu_1 \ldots \mu_m}$) composite operators. The
quantities $C^{NS}_{m,l}$, $C^F_{m,l}$ and $C^{V}_{m,l}$ are called
coefficient functions (CFs) of the OPE.

As a rule, the internal sums in $l$ are omitted in \eqref{OPE} since
neglected terms do not contribute to DIS structure functions (see,
for instance, \cite{Yndurain}). Below we will discuss this point in
more details.

The standard approach to calculations of the OPE CFs is to apply for
perturbation theory by considering the scattering of leptons off
\emph{elementary} (quark and gluon) \emph{off-shell} fields. In
Ref.~\cite{Nonsinglet_case} a new method for calculating CFs was
proposed which does not explicitly depend on elementary fields, but
instead defines the CFs entirely in terms of Green functions of the
currents and/or composite operators.

In our previous paper \cite{Nonsinglet_case} the \emph{nonsinglet}
case was studied. In the present paper we generalize our results for
the \emph{singlet case}. In Section~\ref{sec:OPE} we derive a closed
representation for the singlet CFs in terms of vacuum matrix
elements of the composite operators. In Section~\ref{sec:CF_QCD} we
calculate the singlet CFs in perturbative QCD and demonstrate that
our main formulae not only reproduces well-known expression for the
gluon CF, but enables us to obtain the CFs of all gradient singlet
operators in the OPE. The renormalization of singlet quark and gluon
composite operators and their CFs is considered in
Section~\ref{sec:renorm}. A number of useful mathematical formulae
is collected in Appendix~A (integrals) and Appendix~B (sums).


\section{OPE coefficient functions and vacuum matrix
elements of composite operators}
\label{sec:OPE}

The cross section of deep inelastic lepton-nucleon scattering (DIS)
is related with the hadronic tensor (see, for instance,
\cite{Yndurain})
\begin{equation}\label{hadronic_tensor}
W_{\mu\nu} (p,\, q) = 2\pi^2 \!\! \int \!\! d^4x \,\mbox{e}^{iqx}
\langle \, p \, |\mathrm{T} J^{\mathrm{em}}_{\mu}(x)
J^{\mathrm{em}}_{\nu}(0) | \, p \, \rangle \, .
\end{equation}
Here $| \, p \, \rangle$ means a nucleon state, and
$J^{\mathrm{em}}_{\mu}(x)$ is an electromagnetic current:
\begin{equation}\label{em_current}
J^{\mathrm{em}}_{\mu}(x) =  \overline \Psi (x) \gamma_{\mu} \hat{Q
}\Psi (x) \, ,
\end{equation}
where $ \Psi (x)$ is a quark field. The electric charge operator in
\eqref{em_current},
\begin{equation}\label{electric_charge}
\hat{Q} = \frac{1}{2}(\lambda^3 + \frac{1}{\sqrt{3}} \lambda^8) \, ,
\end{equation}
obeys the equation
\begin{equation}\label{electric_charge_square}
\hat{Q}^2 =  \frac{1}{6} \left( \lambda^3 + \frac{1}{\sqrt{3}}
\lambda^8 \right) + \frac{2}{9} \lambda^0 \, ,
\end{equation}
were $\lambda^a$ ($a = 1,2, \ldots 8$) are the Gell-Mann matrices,
$\mbox{Sp}(\lambda^a) = 0$, $\mbox{Sp}(\lambda^a \lambda^b) = 2
\delta_{ab}$, and $\lambda^0$ is the identity matrix.

For DIS of a charged lepton, the hadronic tensor
\eqref{hadronic_tensor} has two independent tensor structures
\cite{Yndurain}:
\begin{align}\label{tensor structure}
W_{\mu\nu} (p,\, q) & = \left( - g_{\mu\nu} + \frac{q_{\mu}
q_{\nu}}{q^2} \right) \frac{1}{2x_B} \, F_1 (x_B, Q^2)
\nonumber \\
& + \left(p_{\mu} - q_{\mu} \frac{pq}{q^2} \right) \left(p_{\nu} -
q_{\nu} \frac{pq}{q^2} \right) \frac{2 x_B}{Q^2} \, F_2(x_B, Q^2) \,
,
\end{align}
where $Q^2 = - q^2$,
\begin{equation}\label{Bjorken_variable}
x_B = Q^2/2pq
\end{equation}
is the Bjorken variable, and structure functions $F_1$, $F_2$ depend
on these invariant variables. In the Bjorken limit, $Q^2 \rightarrow
\infty$, $x_B$ is fixed, the structure functions $F_{1,2}(Q^2, \,
x)$ are defined via one-nucleon matrix elements of the composite
operators which enter OPE \eqref{OPE}.

\begin{figure}[ht]
\begin{center}
\resizebox{\textwidth}{!}{\includegraphics{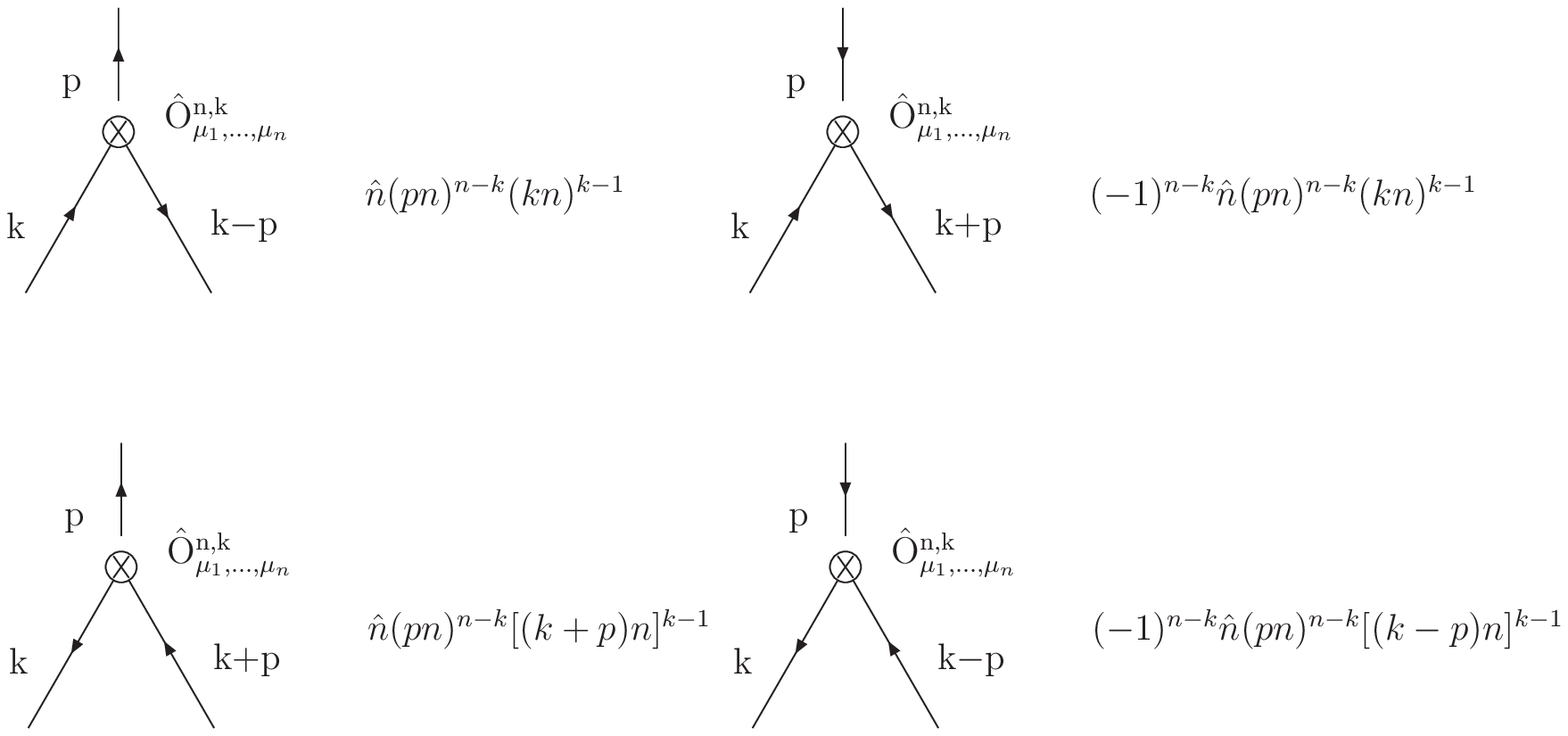}}
\caption{Feynman rules for the quark composite operators
$O^{n,k}_{\mu_1 \ldots \mu_n}$ in the leading (zero) order in strong
coupling $\alpha_s$.} \label{fig:operator_quark_0}
\end{center}
\end{figure}

\begin{figure}[ht]
\begin{center}
\resizebox{\textwidth}{!}{\includegraphics{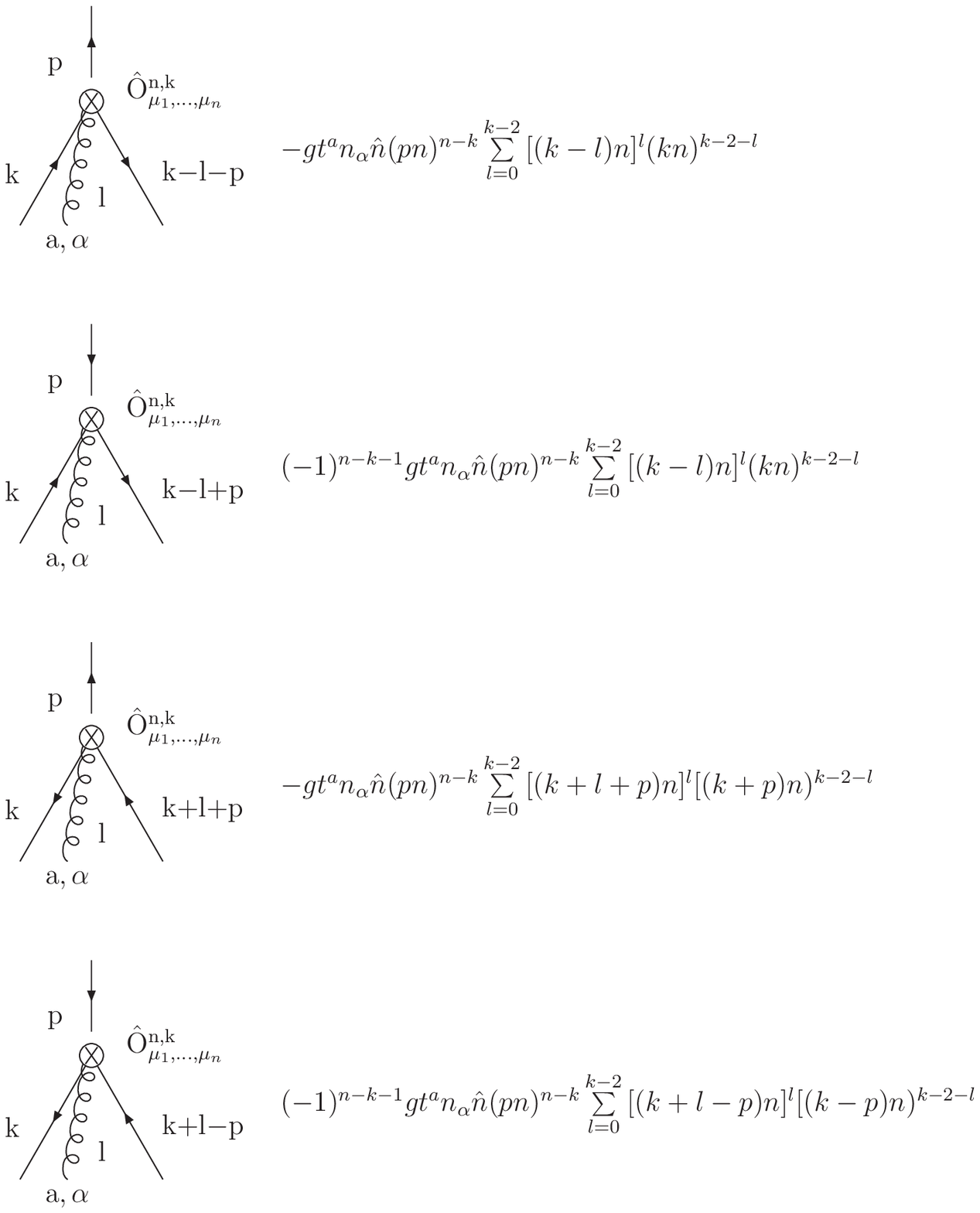}}
\caption{Feynman rules for the quark composite operators
$O^{n,k}_{\mu_1 \ldots \mu_n}$  in the first order in strong
coupling $\alpha_s$.} \label{fig:operator_quark_1}
\end{center}
\end{figure}

Near the light-cone, leading contributions to matrix elements come
from twist-2 operators. In QCD, the nonsinglet quark twist-2
(traceless) gauge-invariant operators%
\footnote{Non-gauge-invariant composite operators in the OPE will be
discussed in the end of section~\ref{sec:renorm}.}
are of the form ($1 \leqslant l \leqslant m$):
\begin{align}\label{nonsinglet_CO}
O^{a,\,m,l}_{NS,\,\mu_1 \ldots \mu_m}(x) &= i^{m-1} \mathbf{S} \,
\partial_{\mu_{l+1}} \ldots
\partial_{\mu_m} \overline \Psi (x) \gamma_{\mu_1} D_{\mu_2}
\ldots D_{\mu_l} \lambda^a \Psi (x)
\nonumber \\
&+ (\mbox{terms proportional to } g_{\mu_i\mu_j}) \, ,
\end{align}
where operator $\mathbf{S}$ means a complete symmetrization in
Lorentz indices,
\begin{equation}\label{14}
D_{\mu} = \partial_{\mu} + ig \, t_a A^a_{\mu}
\end{equation}
is a covariant derivative, and  $A^a_{\mu}(x)$ is a gluon field. The
singlet quark twist-2 operators ($1 \leqslant l \leqslant m$),
\begin{align}\label{quark_singlet_CO}
O^{m,l}_{F, \,\mu_1 \ldots \mu_m}(x) &= i^{m-1} \mathbf{S} \,
\partial_{\mu_{l+1}} \ldots
\partial_{\mu_m} \overline \Psi (x) \gamma_{\mu_1} D_{\mu_2}
\ldots D_{\mu_l} \Psi (x)
\nonumber \\
&+ (\mbox{terms proportional to } g_{\mu_i\mu_j}) \, ,
\end{align}
can mix with the gluon twist-2 operators ($1 \leqslant l \leqslant
m-1$)
\begin{align}\label{gluon_CO}
O^{m,l}_{V, \,\mu_1 \ldots \mu_m}(x) &= i^{m-2} \mathbf{S}
\,\mathrm{Sp} \, \partial_{\mu_{l+1}} \ldots
\partial_{\mu_{m-1}}  F_{\mu_1 \, \alpha}(x) D_{\mu_2}
\ldots D_{\mu_l}  F_{\mu_n}^{\alpha} (x)
\nonumber \\
&+ (\mbox{terms proportional to } g_{\mu_i\mu_j}) \, .
\end{align}
Feynman rules for these composite operators which will be used for
our further calculations are presented in
Figs.~\ref{fig:operator_quark_0}--\ref{fig:operator_quark_1}. They
has to be considered as a generalization of well-known Feynman
rules~\cite{Floratos:77} for the case $p \neq 0$.

If the OPE \eqref{OPE} is applied to DIS, only operators of the type
\begin{align}
O^{a,\,m}_{NS,\,\mu_1 \ldots \mu_m}(x) &= i^{m-1} \mathbf{S} \,
\overline \Psi (x) \gamma_{\mu_1} D_{\mu_2} \ldots D_{\mu_m}
\lambda^a \Psi (x)
\nonumber \\
& + (\mbox{terms proportional to } g_{\mu_i\mu_j})  \, ,
\label{nonsinglet_CO_major} \\
O^{m}_{F,\,\mu_1 \ldots \mu_m}(x) & = i^{m-1} \mathbf{S} \,
\overline \Psi (x) \gamma_{\mu_1} D_{\mu_2} \ldots D_{\mu_m} \Psi
(x)
\nonumber \\
& + (\mbox{terms proportional to } g_{\mu_i\mu_j}) \, ,
\label{quark_singlet_CO_major} \\
O^{m}_{V, \,\mu_1 \ldots \mu_m}(x) &= i^{m-2} \mathbf{S}
\,\mathrm{Sp} \, F_{\mu_1 \, \alpha}(x) D_{\mu_2} \ldots
D_{\mu_{m-1}} F_{\mu_m}^{\alpha} (x)
\nonumber \\
& + (\mbox{terms proportional to } g_{\mu_i\mu_j})
\label{gluon_CO_major}
\end{align}
are relevant. It is due to the fact that any composite operator
$O^{m, \l}_{A, \, \mu_1 \ldots \mu_m}$ ($A = NS, \, F, \, V$) with
at least one full derivative%
\footnote{As one can see from Eqs.~\eqref{nonsinglet_CO},
\eqref{quark_singlet_CO}, \eqref{gluon_CO}, the total number of full
derivatives is equal to $m-l$  or $m-l-1$ for the quark or gluon
composite operator, respectively.}
gives no contribution to a forward matrix element $\langle p | \,
O^{m, \, l}_{A,\,\mu_1 \ldots \mu_m} | p \rangle$. In our notation,
\begin{align}\label{CO_major}
O^{a,\,m}_{NS,\,\mu_1 \ldots \mu_m} & = O^{a,\,m,m}_{NS,\,\mu_1
\ldots \mu_m} \, ,
\nonumber \\
O^{m}_{F,\,\mu_1 \ldots \mu_m} & = O^{m, \, m}_{F,\,\mu_1 \ldots
\mu_m},
\nonumber \\
O^{m}_{V,\,\mu_1 \ldots \mu_m} & = O^{m, \, m-1}_{V,\,\mu_1 \ldots
\mu_m} \, .
\end{align}
In what follow, the operators \eqref{CO_major} will be called
``major'' or ``diagonal'' composite operators, while the quark
operators with $1 \leqslant l \leqslant m-1$ and gluon operators
with $1 \leqslant l \leqslant m-2$ will be reffered to as
``nondiagonal'' composite operators. Correspondingly, we define:
\begin{align}\label{CFs_major}
C_{m}^{NS} & = C_{m,m}^{NS} \, ,
\nonumber \\
C_{m}^{F} & = C_{m,m}^{F} \, ,
\nonumber \\
C_{m}^{V} & = C_{m,m-1}^{V} \, .
\end{align}

For nonforward matrix elements (for instance, describing deeply
virtual Compton scattering), all composite operators contribute.
Namely, we have the relations:%
\footnote{Here $\langle p \, |$ and $| p + \Delta \rangle$ are
one-particle states with 4-momenta $p_{\mu}$ and $ (p +
\Delta)_{\mu}$.}
\begin{align}\label{nonforward_ME}
\langle p + \Delta | \, O^{a, \, m, \l}_{NS, \, \mu_1 \ldots \mu_m}
| p \rangle & = \Delta_{\mu_{l+1}} \ldots \Delta_{\mu_m} \, \langle
p + \Delta|\,  O^{\, a, \, l}_{NS, \, \mu_1 \ldots \mu_l} | p
\rangle \, ,
\nonumber \\
\langle p + \Delta|\,  O^{m, \l}_{F, \, \mu_1 \ldots \mu_m} | p
\rangle & = \Delta_{\mu_{l+1}} \ldots \Delta_{\mu_m} \, \langle p +
\Delta | \, O^{\, l}_{F, \, \mu_1 \ldots \mu_l} | p \rangle  \, ,
\nonumber \\
\langle p + \Delta| \, O^{m, \l}_{V, \, \mu_1 \ldots \mu_m} | p
\rangle  & = \Delta_{\mu_{l+1}} \ldots \Delta_{\mu_{m-1}} \, \langle
p + \Delta | \, O^{\, l}_{V, \, \mu_1 \ldots \mu_l} | p \rangle  \,
.
\end{align}
The ``major'' operators in the RHS of Eq.~\eqref{nonforward_ME} are
defined above in Eq.~\eqref{CO_major}.

As usual, we assume that $C^A_{m,l}(x^2)$ are tempered generalized
functions (this is explicit in perturbative calculations), so the
symbolic relation
\begin{equation}\label{}
x^{\mu_1} \ldots x^{\mu_m} = (-2i)^m \, \frac{q^{\mu_1} \ldots
q^{\mu_m}}{(-q^2)^m} \, (-q^2)^m \left( \frac{\partial}{\partial
q^2} \right)^m
\end{equation}
holds in connection with the Fourier transform in \eqref{OPE}.

The approach, developed in our previous paper \cite{Nonsinglet_case}
for the nonsinglet CFs, should be generalized for the singlet case.
To do this, let us
\begin{enumerate}
  \item take T-product of both sides of the OPE \eqref{OPE} by a
  singlet composite operator $O^{n,k}_{A, \, \nu_1 \ldots
  \nu_n}(z)$, with $A=F \mathrm{\ or \ } V$,
  \item imbed all resulting operator products between vacuum states.
\end{enumerate}
As a result, we obtain from Eq.~\eqref{OPE} the following relation
between vacuum matrix elements of the operator products and OPE
coefficient functions:
\begin{align}\label{JJO_expansion}
\int \! d^4x \,\mbox{e}^{iqx} & \!\! \int \!\! d^4z \,\mbox{e}^{ipz}
\langle \mbox{T}\tilde{ J}^{\mathrm{em}}_{\mu}(x)
\tilde{J}^{\mathrm{em}}_{\nu}(0) \, O^{n,k}_{A, \, \nu_1 \ldots
\nu_n}(z) \rangle
\nonumber \\
& = - g_{\mu\nu} \bigg\{ \sum\limits_{m=0}^{\infty}
\sum\limits_{l=1}^m 2^m \, \frac{q^{\mu_1} \ldots
q^{\mu_m}}{(-q^2)^m} \, \tilde{C}_{m,l}^F(q^2)
\nonumber \\
& \times \, \! \int \! d^4z \,\mbox{e}^{ipz} \langle \mbox{T}
O^{m,l}_{F, \, \mu_1 \ldots \mu_m}(0) \, O^{n,k}_{A, \, \nu_1 \ldots
\nu_n}(z) \rangle
\nonumber \\
& + \sum\limits_{m=0}^{\infty} \sum\limits_{l=1}^{m-1} 2^m \,
\frac{q^{\mu_1} \ldots q^{\mu_m}}{(-q^2)^m} \,
\tilde{C}_{m,l}^V(q^2)
\nonumber \\
& \times \, \! \int \! d^4z \,\mbox{e}^{ipz} \langle \mbox{T}
O^{m,l}_{V, \, \mu_1 \ldots \mu_m}(0) \, O^{n,k}_{A, \, \nu_1 \ldots
\nu_n}(z) \rangle + \ldots \, ,
\end{align}
where $\tilde{C}_{m,l}^A(q^2)$ is a Fourier transform of
$C_{m,l}^A(x^2)$,
\begin{equation}\label{Fourier_CF}
\tilde{C}^A_{m,l}(q^2) = \frac{1}{m!} \, (-q^2)^m \left(
\frac{\partial}{\partial q^2} \right)^m \int \! d^4x
\,\mbox{e}^{iqx} C^A_{m,l}(x^2) \, ,
\end{equation}
and a new notation,
\begin{equation}\label{J}
\tilde{J}^{\mathrm{em}}_{\mu}(x) =  \overline \Psi (x) \gamma_{\mu}
\lambda^0 \Psi (x) \, ,
\end{equation}
is introduced. In other words, only a \emph{singlet part} of the
product of two electromagnetic currents (see
Eqs.~\eqref{em_current}, \eqref{electric_charge_square}) gives a
contribution to \eqref{JJO_expansion}.

Let $n_\mu$ be a light-cone 4-vector not orthogonal to 4-momentum
$p_{\mu}$:
\begin{equation}\label{LC_vector}
n_\mu^2 = 0 \, , \qquad p \, n \neq 0 \, .
\end{equation}
Throughout the paper, we will work in the limit
\begin{equation}\label{p2_limit}
p \rightarrow 0 \, , \qquad p^2 < 0 \, .
\end{equation}
Let us underline that the limit $p^2 \rightarrow 0$ does not assume
the limit $p_{\mu} \rightarrow 0$. On the contrary, given $n_{\mu} =
q_{\mu} - p_{\mu} \, q^2\!/[pq + \sqrt{(pq)^2 - p^2q^2}]$, one gets
$pn \simeq pq$ at $p^2 \simeq 0$.

It is useful to convolute vacuum matrix elements with the projector
\begin{equation}\label{projector}
\frac{n^{\nu_1} \ldots n^{\nu_n}}{(pn)^n} \, ,
\end{equation}
and define the following invariant structure,
\begin{align}\label{JJO_invariant_part}
& \frac{n^{\nu_1} \ldots n^{\nu_n}}{(pn)^n} \int \! d^4x
\,\mbox{e}^{iqx} \int \! d^4z \,\mbox{e}^{ipz} \langle \mbox{T}
\tilde{J}^{em}_{\mu}(x) \tilde{J}^{em}_{\nu}(0) \, O^{n,k}_{A, \,
\nu_1 \ldots \nu_n}(z) \rangle
\nonumber \\
& = - \frac{4}{9} \, g_{\mu\nu} F^{n,k}_A(\omega, Q^2, p^2) + \ldots
\, .
\end{align}
It depends on invariant variables $p^2$,  $Q^2$ and dimensionless
variable
\begin{equation}\label{42}
\omega = 1/x_B = 2pq/Q^2 \, .
\end{equation}
The vacuum matrix element of the T-product of two composite
operators has the following Lorentz structure ($A, \, B = F, \, V$):
\begin{align}\label{matrix_element_OO_1}
\int \! d^4z \,\mbox{e}^{ipz} & \langle \mbox{T} O^{m,l}_{B, \,
\mu_1 \ldots \mu_m}(0) \, O^{n,k}_{A, \, \nu_1 \ldots \nu_n}(z)
\rangle
\nonumber \\
& = 2 \, p_{\mu_1} \ldots p_{\mu_m} p_{\nu_1} \ldots p_{\nu_n}
\langle O^{m,l}_B O^{n,k}_A \rangle (p^2)
\nonumber \\
&  + \, (\mbox{terms proportional to } g_{\mu_i\mu_j}p^2, \
g_{\nu_i\nu_j}p^2, \ g_{\mu_i\nu_j}p^2) \, .
\end{align}
Equation \eqref{matrix_element_OO_1} means:
\begin{align}\label{matrix_element_OO_2}
\frac{n^{\nu_1} \ldots n^{\nu_n}}{(pn)^n} & \! \int \! d^4z \,
\mbox{e}^{ipz} \langle \mbox{T} \, O^{m,l}_{B,\, \mu_1 \ldots
\mu_m}(0) O^{n,k}_{A, \, \nu_1 \ldots \nu_n}(z) \rangle
\nonumber \\
& = 2 \, p_{\mu_1} \ldots p_{\mu_m} \langle  O^{m,l}_B O^{n,k}_A
\rangle (p^2) \, .
\end{align}
Note that both $F^{n,k}_A(\omega, Q^2, p^2)$ and $\langle  O^{m,l}_B
O^{n,k}_A \rangle(p^2)$ are dimensionless quantities.

Let us note that at $p^2 \rightarrow 0$ vacuum matrix elements of
composite operators of higher twists are suppressed by powers of
$p^2$ with respect to the vacuum matrix elements of twist-2
operators \eqref{matrix_element_OO_2}. Thus, our approach enables us
to isolate a contribution from twist-2 operators.

At fixed $Q^2$ and $p^2$, 3-point Green function $\langle \mbox{T}
J^{em}_{\mu} J^{em}_{\nu} \, O^{n,k}_A \rangle$ has a discontinuity
in the variable $(q+p)^2$ for $(q+p)^2 \geqslant 0$ (that is, for
$\omega \geqslant 1$). By using the dispersion relation for
$F^{n,k}_A(\omega, Q^2, p^2)$,
\begin{eqnarray}\label{dispersion_relation}
F^{n,k}_A(\omega, Q^2, p^2) &=& \frac{1}{\pi} \int\limits_1^{\infty}
\! \frac{d \omega'}{\omega' - \omega} \, \mathrm{\rm Im}
F^{n,k}_A(\omega', q^2, p^2)
\nonumber \\
&=& \frac{1}{2\pi i}  \sum\limits_{m=0}^{\infty} \omega^m
\int\limits_1^{\infty} \! d \omega' \omega'^{-m-1} \,
\mathrm{disc}_{\omega} F^{n,k}_A(\omega', q^2, p^2),
\end{eqnarray}
one can derive from Eqs.~\eqref{JJO_expansion} and
\eqref{JJO_invariant_part}, \eqref{matrix_element_OO_2}:
\begin{align}\label{CF_vs_JJO}
& \Bigg[ \sum\limits_{l=1}^m \tilde{C}^F_{m,l}(Q^2/\mu^2) \, \langle
O_F^{m,l} O_A^{n,k} \rangle (p^2/\mu^2)
\nonumber \\
& + \sum\limits_{l=1}^{m-1} \tilde{C}^V_{m,l}(Q^2/\mu^2) \, \langle
O_V^{m,l} O_A^{n,k} \rangle (p^2/\mu^2) \Bigg]_{p^2 \rightarrow 0}
\nonumber \\
& = \left[ \frac{1}{2\pi i} \! \int\limits_0^{1} \! d x_{B}
x_{B}^{m-1} \mbox{disc}_{(p+q)^2} F^{n,k}_A(x_B, Q^2/p^2, p^2/\mu^2)
\right]_{p^2 \rightarrow 0} \, .
\end{align}
Strictly speaking, possible divergencies must be subtracted from the
dispersion relation \eqref{dispersion_relation}. However, it does
not alter our scheme provided the integrals in the r.h.s. of
Eq.~\eqref{CF_vs_JJO} converge (remember that $m \geq 2$).

In \eqref{CF_vs_JJO} we took into account that both matrix elements
of renormalized composite operators and CFs depend on the
\emph{renormalization} scale $\mu$. In what follows, we take $\mu$
to be equal to the \emph{regularization} scale $\bar{\mu}$, which
arises in dimensional regularization \cite{Hooft:72}, when one
changes an integration volume, $d^4\!k \rightarrow \bar{\mu}^{(4-D)}
d^D\!k$.

Both sides of Eq.~\eqref{CF_vs_JJO} has no dependence on $n$ except
for the trivial factor $(-1)^n$. By setting $k = 1, \, 2, \ldots
2m-1$, we thus obtain a set of $2m-1$ algebraic equations for the
singlet OPE CFs $\tilde{C}^F_{m,l}$ ($1 \leq l \leq m$) and
$\tilde{C}^V_{m,l}$ ($1 \leq l \leq m-1 $)\,.%
\footnote{Since $n \geq k$, the index $n$ must be chosen larger than
$2m-1$.}

Formula \eqref{CF_vs_JJO} gives an operator definition of the OPE
CFs in term of vacuum matrix elements of composite operators.%
\footnote{The electromagnetic current \eqref{em_current} is a
particular case of a quark composite operator with zero anomalous
dimension.}
It is important to stress that our definition of the OPE CFs is
unambiguous and it does not lean on a notion of quark and gluon
distributions. The latter are defined via nucleon matrix elements of
the quark or gluon composite operator, while the coefficient
functions are independently defined via vacuum matrix elements of
the product of composite operators.


\section{Calculations of singlet coefficient functions in
perturbative QCD}
\label{sec:CF_QCD}

The formula~\eqref{CF_vs_JJO} is a generalization of a corresponding
formula for a nonsiglet case which was derived in our previous paper
\cite{Nonsinglet_case}:
\begin{align}\label{CF_singlet_vs_JJO}
& \left[ \sum\limits_{l=1}^m \tilde{C}^{NS}_{m,l}(Q^2/\mu^2) \,
\langle O_{NS}^{m,l} \, O_{NS}^{n,k} \rangle (p^2/\mu^2)
\right]_{p^2 \rightarrow 0}
\nonumber \\
& = \left[ \frac{1}{2\pi i} \! \int\limits_0^{1} \! d x_{B} \,
x_{B}^{m-1} \mbox{disc}_{(p+q)^2} F^{n,k}_{NS}(x_B, Q^2/p^2,
p^2/\mu^2) \right]_{p^2 \rightarrow 0} \, .
\end{align}
By using this formula, the following expressions for the nonsinglet
CFs were calculated in QCD \cite{Nonsinglet_case}:
\begin{align}
\left[ \tilde{C}^{NS}_{m,m} \right]^{(0)} & = \frac{1}{2} \, [1 +
(-1)^m],
\label{CF_NS_0_main} \\
\intertext{and}
\left[ \tilde{C}^{NS}_{m,l} \right]^{(0)} & = \frac{1}{2} \,
(-1)^l \, {m\choose{l}} \, ,
\label{CF_NS_0_rest}
\end{align}
for $l=0,1, \ldots, m-1$.%
\footnote{Everywhere ${n\choose{m}}$ denotes a binomial
coefficient.}
Here and in what follows superscript ``(0)'' means that a
corresponding quantity is calculated in zero order in strong
coupling.

In the next order in  $\alpha_s$, we obtained the following
expressions \cite{Nonsinglet_case}:
\begin{align}
& \left[ \tilde{C}^{NS}_{m,m} (Q^2/\mu^2) \right]^{(1)} =
\frac{\alpha_s}{8 \pi} \, C_F \ln \left( \frac{Q^2}{\mu^2} \right)
\,
\nonumber \\
& \times [1 + (-1)^m] \left[ -4 \sum_{j=2}^m \frac{1}{j} - 1 +
\frac{2}{m(m+1)} \right] \, ,
\label{CF_NS_1_main} \\
\intertext{and} & \left[ \tilde{C}^{NS}_{m,l} (Q^2/\mu^2)
\right]^{(1)}=  \frac{\alpha_s}{4 \pi} \, C_F \ln \left(
\frac{Q^2}{\mu^2} \right)
\nonumber \\
& \times \bigg\{ \frac{1}{2} (-1)^l {m-1\choose{l-1}} \left[ -4
\sum_{j=2}^l  \frac{1}{j}  - 1 + \frac{2}{l(l+1)} \right]
\nonumber \\
& + \left( \frac{1}{m-l} - \frac{1}{m+1} \right)
\nonumber \\
& + \sum_{k=l+1}^m (-1)^k {m-1\choose{k-1}} \left( \frac{1}{k-l} -
\frac{1}{k+1} \right) \bigg\} \, ,
\label{1CF_NS_1_rest}
\end{align}
for $l=0,1, \ldots, m-1$. Here and in what follows superscript
``(1)'' means that a corresponding quantity is calculated in the
first order in strong coupling constant.

Let us stress that we did not demand  from the very beginning that
the ``major'' CF, $\tilde{C}^{NS}_{m,m}$, should be equal to zero
for odd $m$ (see Eqs.~\eqref{CF_NS_0_main} and
\eqref{CF_NS_1_main}). On the contrary, it is a consequence of the
fact that electromagnetic interactions conserve P-parity. Remember
that DIS structure function $F_2(x_B, Q^2)$ is an even function of
Bjorken variable $x_B$, and its nonzero moments, $F_2(n, Q^2)$, are
defined by quantities $\tilde{C}^{A}_{n,n} \, (Q^2/\mu^2) \langle
\,p\,|\, \hat{O_A^{n,n}}\,|\,p \,\rangle (\mu^2)$, with even $n$.
That is why we expect that the gluon ``major'' CF should be also
proportional to the factor $[1 + (-1)^m]$ (see \eqref{gluon_CF_main}
below).

For a convenience, let us for a moment rewrite our main relation
\eqref{CF_singlet_vs_JJO} in simbolic form:
\begin{equation}\label{CF_vs_JJO_simple}
\langle J J \, O_A \rangle = C^F \langle O_F \, O_A \rangle + C^V
\langle O_V \, O_A \rangle \, .
\end{equation}
Then we get from \eqref{CF_vs_JJO_simple}:
\begin{equation}\label{OV_vs_JJO__0}
\left[ \langle J J \, O_V \rangle \right]^{(0)}  = \left[ C^F
\right]^{(0)} \left[ \langle O_F \, O_V \rangle \right]^{(0)} +
\left[ C^V \right]^{(0)} \left \langle O_V \, O_V \rangle
\right]^{(0)} \, .
\end{equation}
Since $[ \langle O_V \, O_F \rangle ]^{(0)} = [ \langle J J \, O_V
\rangle ]^{(0)} = 0$, while $[\langle O_V \, O_V \rangle ]^{(0)}$ is
nonzero, we get:
\begin{equation}\label{CF_gluon_0_1}
\sum\limits_{l=1}^{m-1} \left[ \tilde{C}^V_{m,l} \right]^{(0)}
\left[\langle O_V^{m,l} O_V^{n,k} \rangle \right]^{(0)}= 0 \, .
\end{equation}
Equality \eqref{CF_gluon_0_1} is valid for all integer $m$, $n$ and
$1 \leqslant k \leqslant n-1$. Thus, we conclude that
\begin{equation}\label{CF_gluon_0_2}
\left[ \tilde{C}^V_{m,l} \right]^{(0)} = 0 \, ,
\end{equation}
for all integer $m$, and $1 \leqslant l \leqslant m-1$.

Analogously, we obtain from \eqref{CF_vs_JJO_simple}:
\begin{align}
\left[ \langle J J \, O_F \rangle \right]^{(0)} & = \left[ C^F
\right]^{(0)} \left[ \langle O_F \, O_F \rangle \right]^{(0)} +
\left[ C^V \right]^{(0)} \left \langle O_V \, O_F \rangle
\right]^{(0)} \, ,
\label{OF_vs_JJO__0} \\
\left[ \langle J J \, O_F \rangle \right]^{(1)} & = \left[ C^F
\right]^{(0)} \left[ \langle O_F \, O_F \rangle \right]^{(1)} +
\left[ C^F \right]^{(1)} \left \langle O_F \, O_F \rangle
\right]^{(0)}
\nonumber \\
& + \left[ C^V \right]^{(0)} \left[ \langle O_V \, O_F \rangle
\right]^{(1)} + \left[ C^V \right]^{(1)} \left \langle O_V \, O_F
\rangle \right]^{(0)} \, . \label{OF_vs_JJO__1}
\end{align}
Taking into account that $[ C^V]^{(0)} = 0$ \eqref{CF_gluon_0_2} and
$[ \langle O_V \, O_F \rangle ]^{(0)} = 0$, we find:
\begin{align}
\left[ \langle J J \, O_F \rangle \right]^{(0)} & = \left[ C^F
\right]^{(0)} \left[ \langle O_F \, O_F \rangle \right]^{(0)} \, ,
\label{OF_vs_JJO__0} \\
\left[ \langle J J \, O_F \rangle \right]^{(1)} & = \left[ C^F
\right]^{(0)} \left[ \langle O_F \, O_F \rangle \right]^{(1)} +
\left[ C^F \right]^{(1)} \left \langle O_F \, O_F \rangle
\right]^{(0)} \, .
\label{OF_vs_JJO__1}
\end{align}
These equations are identical to those derived for the
\emph{nonsinglet} quark CF in our paper \cite{Nonsinglet_case}. As a
result, we find that the singlet quark CFs coincide with the
corresponding nonsinglet quark CFs in zero and first order in
$\alpha_s$:%
\footnote{Note, however, that $[\tilde{C}^F_{m,l}]^{(n)} \neq
[\tilde{C}^{NS}_{m,l} ]^{(n)}$, for $n \geqslant 2$.}
\begin{align}\label{CF_quark_nonsinglet}
\left[ \tilde{C}^F_{m,l} \right]^{(0)} &  = \left[
\tilde{C}^{NS}_{m,l} \right]^{(0)} \, ,
\nonumber \\
\left[ \tilde{C}^F_{m,l} \right]^{(1)} &  = \left[
\tilde{C}^{NS}_{m,l} \right]^{(1)} \, ,
\end{align}
with $[\tilde{C}^{NS}_{m,l}]^{(0)}$ and
$[\tilde{C}^{NS}_{m,l}]^{(1)}$ given by expressions
\eqref{CF_NS_0_main}--\eqref{1CF_NS_1_rest}.

Now let us turn to QCD calculations of the gluon CFs in the first
order in strong coupling constant by using our main
formula~\eqref{CF_vs_JJO}. We work in the dimensional regularization
\cite{Hooft:72} and use the $\overline{\mathrm{MS}}$-scheme to
renormalize ultraviolet divergences. All results of our calculations
are gauge invariant since we sum all diagrams in each order of
perturbation theory. Let us remember that in order to find the OPE
CFs, we have to retain only leading terms in the limit $p^2
\rightarrow 0$. This significantly simplifies the calculations. We
will restrict ourselves by considering leading terms in $\ln
(Q^2/\mu^2)$, although our main formula~\eqref{CF_vs_JJO} enables
one to calculate subleading terms as well. In other words, along
with the limit $p^2 \rightarrow 0$, we are interested in large
values of variable $Q^2$.

Starting from \eqref{CF_vs_JJO_simple}, we can schematically write:
\begin{equation}\label{OV_vs_JJO_simple}
\left[ \langle J J \, O_V \rangle \right]^{(1)} = \left[ C^F
\right]^{(0)} \left[ \langle O_F \, O_V \rangle \right]^{(1)} +
\left[ C^V \right]^{(1)} \left \langle O_V \, O_V \rangle
\right]^{(0)} \, .
\end{equation}
In full detail, Eq.~\eqref{OV_vs_JJO_simple} looks like the
following:
\begin{align}\label{CV_vs_JJO_1}
& \left\{ \frac{1}{2\pi i} \! \int\limits_0^{1} \! d x_{B} \,
x_{B}^{m-1} \left[ \mbox{disc}_{(p+q)^2} F^{n,k}_V(x_B, Q^2/p^2,
p^2/\mu^2) \right]^{(1)} \right\}_{p^2 \rightarrow 0}
\nonumber \\
& = \Bigg\{ \sum\limits_{l=1}^m \left[ \tilde{C}^F_{m,l}
\right]^{(0)} \left[ \langle O_F^{m,l} O_V^{n,k} \rangle (p^2/\mu^2)
\right]^{(1)}
\nonumber \\
& \hspace{.5cm} + \sum\limits_{l=1}^{m-1} \left[
\tilde{C}^V_{m,l}(Q^2/\mu^2) \right]^{(1)} \left[ \langle O_V^{m,l}
O_V^{n,k} \rangle (p^2/\mu^2) \right]^{(0)} \Bigg\}_{p^2 \rightarrow
0} \, .
\end{align}
The quantities $[\tilde{C}^F_{m,l}]^{(0)}$ are already known (see
\eqref{CF_quark_nonsinglet}, \eqref{CF_NS_0_main},
\eqref{CF_NS_0_rest}), while the other terms in \eqref{CV_vs_JJO_1}
should be calculated.

\begin{figure}[ht]
\begin{center}
\resizebox{.35\textwidth}{!} {\includegraphics{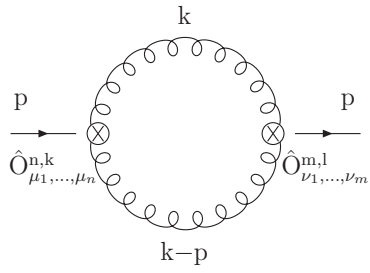}}
\caption{The diagrams for the propagator of the gluon composite
operator $\langle O^{n,k}_V O^{m,l}_V \rangle^{(0)}$ in zero order
in strong coupling $\alpha_s$.} \label{fig:loop_gluon_0}
\end{center}
\end{figure}

The propagator of the gluon composite operator is shown in
Fig.~\ref{fig:loop_gluon_0}, and one gets:%
\footnote{Everywhere $\mathrm{B}(x,y)$ means beta-function.}
\begin{equation}\label{gluon_loop}
\langle O^{n,k}_V O^{m,l}_V \rangle^{(0)} (p^2/\mu^2) =
\mathrm{i}(-1)^{n+1} \frac{1}{8\pi^2} \ln\left(
\frac{\mu^2}{-p^2}\right) \mathrm{B}(k+2,l+2) \, .
\end{equation}

\begin{figure}[ht]
\begin{center}
\resizebox{.8\textwidth}{!}
{\includegraphics{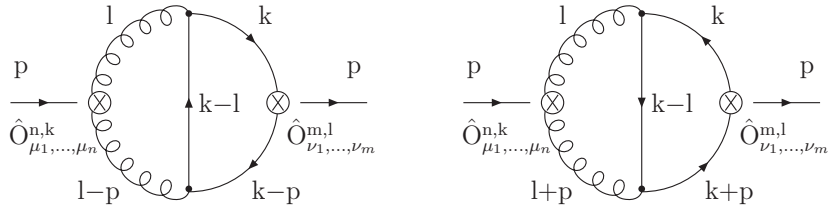}} \caption{The diagrams for
the mixing of the composite operators $\langle O^{n,k}_V O^{m,l}_F
\rangle^{(1)}$ in the first order in strong coupling $\alpha_s$.}
\label{fig:loop_quark-gluon_1}
\end{center}
\end{figure}

The vacuum matrix element of the product of two singlet composite
operators is given by the diagram in
Fig.~\ref{fig:loop_quark-gluon_1}. The calculations result in the
following expression:
\begin{align}\label{quark-gluon_loop}
\langle O^{n,k}_V O^{m,l}_F \rangle^{(1)} (p^2/\mu^2) & = i (-1)^{n}
\, C_F \, \frac{\alpha_s}{4\pi^3} \left[ \ln \left(
\frac{\mu^2}{-p^2}\right) \right]^2
\nonumber \\
& \Bigg\{ \left( \frac{2}{k} - \frac{2}{k+1} + \frac{1}{k+2} \right)
\left[ \frac{1}{k+l+2} - \mathrm{B}(k+3,l) \right]
\nonumber \\
& - \frac{1}{k} \left[ \frac{1}{k+l+1} - \mathrm{B}(k+2,l) \right]
\nonumber \\
& - \left( \frac{1}{k} - \frac{2}{k+1} + \frac{1}{k+2} \right)
\frac{l-1}{l(l+1)} \Bigg\} \, ,
\end{align}
Now we are able to calculate the second term in
Eq.~\eqref{CV_vs_JJO_1}:
\begin{align}\label{quark-gluon_loop_sum}
\sum_{l=1}^{m} & \left[ \tilde{C}^F_{m,l} \right]^{(0)} \langle
O^{n,k}_V O^{m,l}_F \rangle^{(1)} (p^2/\mu^2)
\nonumber \\
& = i (-1)^{n} \,  C_F \, \frac{\alpha_s}{32\pi^3} \left[ \ln \left(
\frac{\mu^2}{-p^2}\right) \right]^2 \frac{1}{m(m+1)(m+2)}
\nonumber \\
& \times   \Bigg\{ \left[ \frac{m-2}{k} + \frac{2(m+2)}{(k+1)(k+2)}
+ \frac{m(m+1)+2}{k+m+2} - \frac{m(m+2)}{k+m+1} \right]
\nonumber \\
& \hspace{.8cm} - \left[ (m-2) \mathrm{B}(k, m+3) + m(m+2)
\mathrm{B}(k+1, m+2) \right] \Bigg\} \, ,
\end{align}
Summation in $l$ was made with the help of
Eqs.~\eqref{Bkl02}-\eqref{Bkl16} from Appendix~B. The relation
between beta-functions (integer $k$, $m$),
\begin{align}\label{beta_func}
&  \left( \frac{2}{k} - \frac{2}{k+1} +\frac{1}{k+2} \right)
\mathrm{B}(k+3, m) -\frac{1}{k} \, \mathrm{B}(k+2, m) = -
\frac{1}{m(m+1)(m+2)}
\nonumber \\
& \times \left[ (m-2) \mathrm{B}(k, m+3) + m(m+2) \mathrm{B}(k+1,
m+2) \right] \, ,
\end{align}
was also used.

\begin{figure}[ht]
\begin{center}
\resizebox{.75\textwidth}{!}{\includegraphics{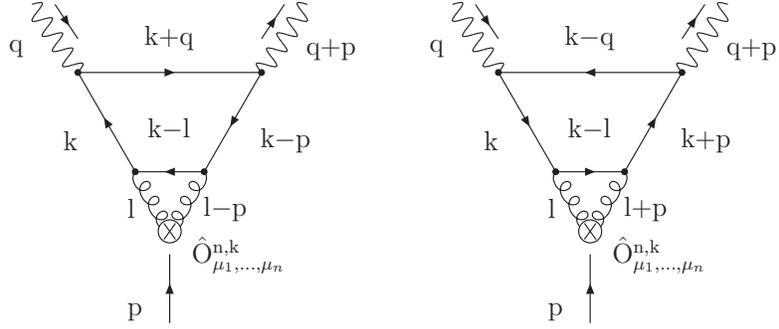}}
\caption{The diagrams  for the matrix element $\langle J J \,
O^{n,k}_V \rangle^{(1)}$ in the first order in strong coupling
$\alpha_s$.} \label{fig:vertex_1}
\end{center}
\end{figure}

The diagrams which contribute to vacuum matrix element with two
currents are shown in Fig.~\ref{fig:vertex_1}. Omitting details of
calculations, let us give the result:
\begin{align}\label{vertex_int}
& \frac{1}{2\pi \mathrm{i}} \int\limits_0^{1} \! d x x^{m-1} \,
\mathrm{disc}_{(p+q)^2} \, \langle J\,J \, O^{n,k}_V \rangle^{(1)}
(x_B, Q^2/p^2, p^2/\mu^2)
\nonumber \\
& = i (-1)^{n} \,  C_F \, \frac{\alpha_s}{32\pi^3} \ln \left(
\frac{Q^2}{-p^2}\right) \ln \left( \frac{\mu^2}{-p^2}\right)
\frac{1}{m(m+1)(m+2)}
\nonumber \\
& \times  \left\{ \left[ \frac{m-2}{k} + \frac{2(m+2)}{(k+1)(k+2)}
\right. + \frac{m(m+1)+2}{k+m+2} - \frac{m(m+2)}{k+m+1} \right]
\nonumber \\
& \hspace{.8cm} - \left[ (m-2) \mathrm{B}(k, m+3) + m(m+2)
\mathrm{B}(k+1, m+2) \right] \bigg\} \, ,
\end{align}
In order to obtain Eqs.~\eqref{gluon_loop}-\eqref{vertex_int}, we
used integrals \eqref{A02}-\eqref{A06} from Appendix~A.

Equations \eqref{CV_vs_JJO_1}-\eqref{vertex_int} result in a set of
equations for the gluon CFs. Namely, for \emph{any} integer $m
\geqslant 2$ we obtain algebraic equations for $[\tilde{C}^V_{m,l}
(Q^2/\mu^2)]^{(1)}$, with $1 \leqslant l \leqslant m-1$:
\begin{align}\label{gluon_CF_equation}
& \sum_{l=1}^{m-1} \left[ \tilde{C}^V_{m,l} (Q^2/\mu^2)
\right]^{(1)} \mathrm{B}(k+2,l+2) = C_F \, \frac{\alpha_s}{4\pi} \ln
\left( \frac{Q^2}{\mu^2}\right) \frac{1}{m(m+1)(m+2)}
\nonumber \\
& \times   \Bigg\{ \left[ \frac{m-2}{k} + \frac{2(m+2)}{(k+1)(k+2)}
+ \frac{m(m+1)+2}{k+m+2} - \frac{m(m+2)}{k+m+1} \right]
\nonumber \\
& \hspace{.8cm} - \left[ (m-2) \mathrm{B}(k, m+3) + m(m+2)
\mathrm{B}(k+1, m+2) \right] \Bigg\}\, .
\end{align}
Note that these equations holds for \emph{all} integer $k \geqslant
1$, but for our purposes it is enough to consider only $m-1$
equations corresponding to $k = 1, 2, \ldots m-1$.%
\footnote{The other equations which correspond to $k \geqslant m$,
will be also satisfied, as one could see from an explicit expression
for our solution \eqref{CF_full}-\eqref{CF_odd}.}

The solution of equations \eqref{gluon_CF_equation} is a sum of two
terms one of which is nonzero only for even $m$, while another is
nonzero only for odd $m$:
\begin{align}\label{CF_full}
\left[ \tilde{C}^V_{m,l}(Q^2/\mu^2) \right]^{(1)} &  = \frac{[1+
(-1)^m]}{2} \left[ \tilde{C}^V_{m,l}(Q^2/\mu^2)
\right]^{(1)}_{\mathrm{even}}
\nonumber \\
& + \frac{[1 - (-1)^m]}{2} \left[ \tilde{C}^V_{m,l}(Q^2/\mu^2)
\right]^{(1)}_{\mathrm{odd}} \, .
\end{align}
The first term in \eqref{CF_full} is defined for $1 \leqslant l
\leqslant m-1$
\begin{align}\label{CF_even}
& \left[ \tilde{C}^V_{m,l}(Q^2/\mu^2) \right]^{(1)}_{\mathrm{even}}
= C_F \, \frac{\alpha_s}{4\pi} \ln \left( \frac{Q^2}{\mu^2} \right)
\nonumber \\
& \times \Bigg\{ \left( \frac{1}{m} - \frac{2}{m+1} + \frac{2}{m+2}
\right) \left[ (-1)^l {m\choose{l+1}} - (m - l) \right]
\nonumber \\
& \hspace{.9cm} - \frac{1}{m+1} \left[ (-1)^l {m-1\choose{l+1}} - (m
- l - 1) \right] \Bigg\} \, ,
\end{align}
while the second term in \eqref{CF_full} is nonzero for $1 \leqslant
l \leqslant m-2$:
\begin{align}\label{CF_odd}
& \left[ \tilde{C}^V_{m,l}(Q^2/\mu^2) \right]^{(1)}_{\mathrm{odd}} =
C_F \, \frac{\alpha_s}{4\pi} \ln\left( \frac{Q^2}{\mu^2} \right) \,
\, \frac{1}{m(m+1)(m+2)}
\nonumber \\
& \times  \bigg\{  (-1)^l \left[  (l - m + 1)(m + 2) + m(m + 1) + 2
\right] {m\choose{l+1}}
\nonumber \\
& \hspace{.8cm} - \left[ (l - m + 1)(m - 2) + m(m + 1) + 2 \right]
\bigg\} \, .
\end{align}
Note that $(\tilde{C}^V_{m,l})_{\mathrm{odd}}$ \eqref{CF_odd} gives
no contribution to $\tilde{C}^V_{m,l}$ \eqref{CF_full} for $l=m-1$
due to relation $[(-1)^m - 1][(-1)^m + 1] = 0$. It is rather easy to
demonstrate that \eqref{CF_full} does obey set of equations
\eqref{gluon_CF_equation} for any $m \geqslant 2$, $k \geqslant 1$,
if one uses formulae \eqref{Bl02}-\eqref{Bl08} from Appendix~B.
Indeed, these formulae lead us to the relations:
\begin{align}\label{zero_loop_sum}
\sum_{l=1}^{m-1} & \left[ \tilde{C}^V_{m,l} (Q^2/\mu^2)
\right]^{(1)}_{\mathrm{even}} \mathrm{B}(k+2,l+2) = \sum_{l=1}^{m-1}
\left[ \tilde{C}^V_{m,l} (Q^2/\mu^2) \right]^{(1)}_{\mathrm{odd}}
\mathrm{B}(k+2,l+2)
\nonumber \\
& =  C_F \, \frac{\alpha_s}{4\pi} \ln \left(
\frac{Q^2}{\mu^2}\right) \frac{1}{m(m+1)(m+2)}
\nonumber \\
& \times   \Bigg\{ \left[ \frac{m-2}{k} + \frac{2(m+2)}{(k+1)(k+2)}
+ \frac{m(m+1)+2}{k+m+2} - \frac{m(m+2)}{k+m+1} \right]
\nonumber \\
& \hspace{.8cm} - \left[ (m-2) \mathrm{B}(k, m+3) + m(m+2)
\mathrm{B}(k+1, m+2) \right] \Bigg\}\, ,
\end{align}

In particular, it follows from Eq.~\eqref{CF_full}, \eqref{CF_even}
that
\begin{align}\label{gluon_CF_main}
\left[ \tilde{C}^V_{m,m-1}(Q^2/\mu^2) \right]^{(1)} & =  C_F \,
\frac{\alpha_s}{4\pi} \ln\left( \frac{Q^2}{\mu^2}\right) \, [1 +
(-1)^m]
\nonumber \\
& \times \left( \frac{1}{m} - \frac{2}{m+1} + \frac{2}{m+2} \right)
\, .
\end{align}
As one can see, ``major'' CF \eqref{gluon_CF_main} is defined by
well-known anomalous dimension~\cite{Gross:73}%
\footnote{In order to obtain the expression for $\gamma_{FV}^n$ in
standard notations, one has to replace index $m$ by $(n-1)$ in
\eqref{anomolous_dimension}.}
\begin{equation}\label{anomolous_dimension}
\gamma_{FV}^m = \frac{\alpha_s}{\pi} \, C_F \, \frac{(m+1)^2 + (m+1)
+ 2}{m(m+1)(m+2)},
\end{equation}

Thus, we have reproduced the standard expression for the``major''
CF, $\tilde{C}^V_{m, m-1}(Q^2/\mu^2)$, and, which is more important,
have calculated ``gradient'' gluon CFs, $\tilde{C}^V_{m,
l}(Q^2/\mu^2)$ ($l = 1, 2. \ldots m-2$), in the first order in
strong coupling $\alpha_s$.


\section{Renormalization of singlet composite
operators and coefficient functions}
\label{sec:renorm}

Let us consider in more detail products of \emph{renormalized}
composite operators  and corresponding \emph{renormalized} CFs which
enter the OPE of two electromagnetic currents~\eqref{OPE}. Both
singlet composite operators, $O^{m,l}_F$ and $O^{m,l}_V$,  depend on
the renormalization scale $\mu_0$, and mix with each other under
rescaling $\mu_0 \rightarrow \mu$:
\begin{align}\label{operator_renorm}
O^{m,k}_{F}(\mu_0^2)  & = \sum_{l=1}^{k} (\hat{Z}^{FF})^k_{l}
(\mu_0^2/\mu^2) \, O^{m,l}_F (\mu^2) + \sum_{l=1}^{k-1}
(\hat{Z}^{FV})^k_{l} (\mu_0^2/\mu^2) \, O^{m,l}_V (\mu^2) ,
\nonumber \\
O^{m,k}_{V}(\mu_0^2)  & = \sum_{l=1}^{k+1} (\hat{Z}^{VF})^k_{l}
(\mu_0^2/\mu^2) \, O^{m,l}_F (\mu^2) + \sum_{l=1}^{k}
(\hat{Z}^{VV})^k_{l} (\mu_0^2/\mu^2) \, O^{m,l}_V (\mu^2) \, ,
\end{align}
where $\mathbf{\hat{Z}}$ are the matrices of a finite
renormalization of the composite operators. In its turn,
\eqref{operator_renorm} means that
\begin{align}\label{CF_renorm_main_1}
\tilde{C}^V_{m,m-1}(Q^2/\mu^2) &= \tilde{C}^F_{m,m}(Q^2/\mu_0^2) \,
(\hat{Z}^{FV})^m_{m-1}(\mu_0^2/\mu^2)
\nonumber \\
& + \tilde{C}^V_{m,m-1}(Q^2/\mu_0^2) \,
(\hat{Z}^{VV})^{m-1}_{m-1}(\mu_0^2/\mu^2) \, ,
\end{align}
and
\begin{align}\label{CF_renorm_rest_1}
\tilde{C}^V_{m,l}(Q^2/\mu^2) & = \sum_{k=l+1}^{m}
\tilde{C}^F_{m,k}(Q^2/\mu_0^2) \,
(\hat{Z}^{FV})^k_{l}(\mu_0^2/\mu^2)
\nonumber \\
& + \sum_{k=l}^{m} \tilde{C}^V_{m,k}(Q^2/\mu_0^2) \,
(\hat{Z}^{VV})^k_{l}(\mu_0^2/\mu^2) \, ,
\end{align}
for $l=1,2, \ldots m-2$.

Since the quantity $\mu_0$ is an arbitrary scale, one can put
$\mu_0^2 = Q^2$ in \eqref{CF_renorm_main_1},
\eqref{CF_renorm_rest_1}, and obtain:
\begin{align}\label{CF_renorm_main_2}
\tilde{C}^V_{m,m-1}(Q^2/\mu^2) &= \tilde{C}^F_{m,m}(1) \,
(\hat{Z}^{FV})^m_{m-1}(Q^2/\mu^2)
\nonumber \\
& + \tilde{C}^V_{m,m-1}(1) \, (\hat{Z}^{VV})^{m-1}_{m-1}(Q^2/\mu^2)
\, , \\
\intertext{and} \tilde{C}^V_{m,l}(Q^2/\mu^2) & = \sum_{k=l+1}^{m}
\tilde{C}^F_{m,k}(1) \, (\hat{Z}^{FV})^k_{l}(Q^2/\mu^2)
\nonumber \\
& + \sum_{k=l}^{m-1} \tilde{C}^V_{m,k}(1) \,
(\hat{Z}^{VV})^k_{l}(Q^2/\mu^2) \label{CF_renorm_rest_2} \, ,
\end{align}
for $l=1,2, \ldots, m-2$. As a result, we find equations for the
\emph{leading} parts of the gluon CFs in the first order in the
strong coupling ($1 \leqslant l \leqslant m-1$):
\begin{align}\label{CF_renorm_rest_3}
\left[ \tilde{C}^V_{m,l}(Q^2/\mu^2) \right]^{(1)}  =
\sum_{k=l+1}^{m} \left[ \tilde{C}^F_{m,k} \right]^{(0)} & \!\left[
(\hat{Z}^{FV})^k_{l}(Q^2/\mu^2) \right]^{(1)} \, \, .
\end{align}
In deriving relation \eqref{CF_renorm_rest_3}, we took into account
that $[ \tilde{C}^V_{m,l}]^{(0)} \!\! = 0$ for all $m$ and $1
\leqslant l \leqslant m-1$ \eqref{CF_gluon_0_2}. By using explicit
form of $[\tilde{C}^F_{m,k}]^{(0)}$ (see
\eqref{CF_quark_nonsinglet}, \eqref{CF_NS_0_main}), these equations
can be written as follows ($1 \leqslant l\leqslant m-1$):
\begin{align}\label{CF_renorm_rest_4}
\left[ \tilde{C}^V_{m,l}(Q^2/\mu^2) \right]^{(1)} & = [1 + (-1)^m]
\left[ (\hat{Z}^{FV})^m_{l}(Q^2/\mu^2) \right]^{(1)}
\nonumber \\
& + \sum_{k=l+1}^{m-1} (-1)^k  {m-1\choose{k-1}} \left[
(\hat{Z}^{FV})^k_{l}(Q^2/\mu^2) \right]^{(1)} \, .
\end{align}
Note that the last term in \eqref{CF_renorm_rest_4} is identically
zero at $l=m-1$.

\begin{figure}[ht]
\begin{center}
\resizebox{.8\textwidth}{!}{\includegraphics{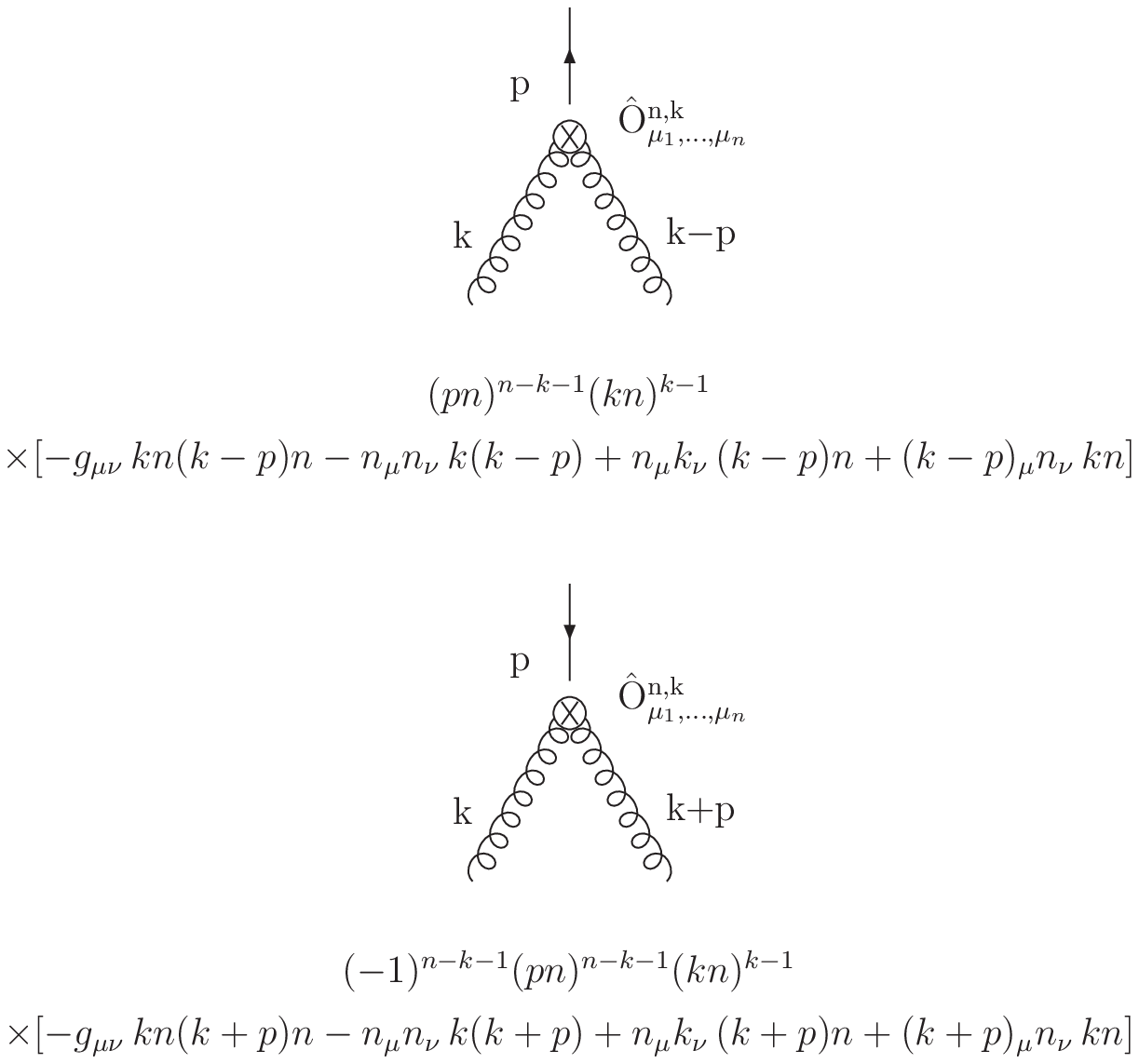}}
\caption{Feynman rules for the gluon composite operators
$O^{n,k}_{\mu_1 \ldots \mu_n}$ in the leading (zero) order in strong
coupling $\alpha_s$.} \label{fig:operator_gluon_0}
\end{center}
\end{figure}

\begin{figure}[ht]
\begin{center}
\resizebox{.7\textwidth}{!}{\includegraphics{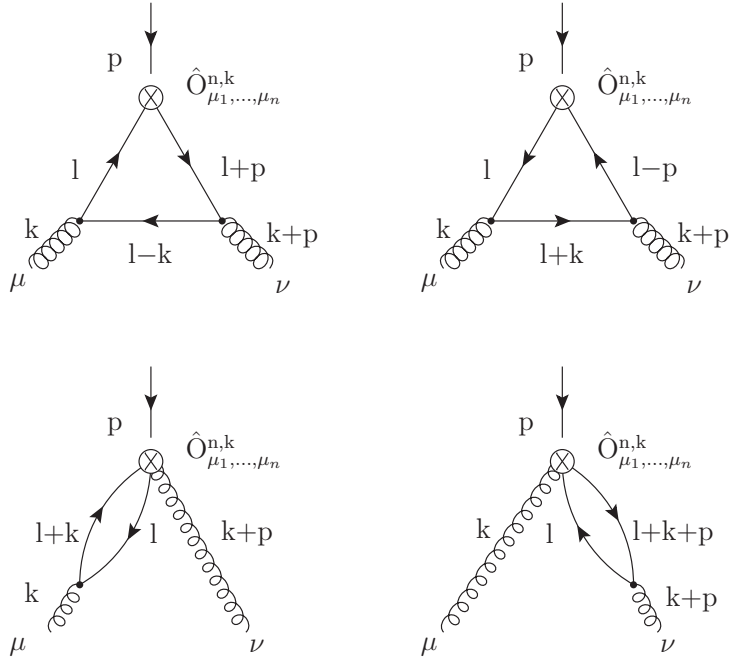}}
\caption{The diagrams which give contribution to the renormalization
of the quark composite operator $O^{n,k}_F$ in the first order in
strong coupling $\alpha_s$.} \label{fig:renorm_quark-gluon}
\end{center}
\end{figure}

The mixing of singlet quark operators \eqref{quark_singlet_CO} and
gluon operators \eqref{gluon_CO} is defined by the set
of diagrams presented in Fig.~\ref{fig:renorm_quark-gluon}.%
\footnote{The expressions for unrenormalized singlet quark and gluon
composite operators are presented in Fig.~\ref{fig:operator_quark_0}
and Fig.~\ref{fig:operator_gluon_0}, respectively.}
The sum of divergent parts of these diagrams is given by the
expression:
\begin{align}\label{Z_FV_diagram}
& \left[ -g_{\mu \nu} \, kn(k+p)n - n_{\mu} n_{\nu} \, k(k+p)
 + n_{\mu}k_{\nu} \, (k+p)n + k_{\mu} n_{\nu} \, kn \right]
\nonumber \\
&  \times C_F \, \frac{\alpha_s}{8\pi} \frac{1}{\varepsilon} \left(
\frac{\mu^2}{-p^2} \right)^{\!\!\varepsilon} \, \sum_{l=0}^{k-1}
(-1)^{n-l-1} (kn)^{l-1} (pn)^{n-l-1}
\nonumber \\
& \times \bigg\{ \left( \frac{1}{k} - \frac{2}{k+1} + \frac{2}{k+2}
\right) (k - l)  - \frac{1}{k+1} (k - l - 1)
\nonumber \\
& - (-1)^l  \left[ \left( \frac{1}{k} - \frac{2}{k+1} +
\frac{2}{k+2} \right) {k\choose{l+1}} -  \frac{1}{k+1}
{k-1\choose{l+1}} \right] \bigg\} \, .
\end{align}
In deriving \eqref{Z_FV_diagram}, basic integrals
\eqref{A12}-\eqref{A18} from Appendix~A were used. Let us remember
Feynman rule for the unrenormalized gluon operator $O^{n,k}_V$ (see
Fig.~\ref{fig:operator_gluon_0}):
\begin{align}\label{operator_quark_unrenorm}
(-1)^{n-k-1} (kn)^{k-1} (pn)^{n-k-1} & \big[ -g_{\mu \nu} \,
kn(k+p)n - n_{\mu} n_{\nu} \, k(k+p)
\nonumber \\
& \hspace{.2cm} + n_{\mu}k_{\nu} \, (k+p)n + k_{\mu} n_{\nu} \, kn
\big] \, .
\end{align}

As a result, the matrix of the finite renormalization in
Eq.~\eqref{CF_renorm_rest_3} has the following form ($1 \leqslant l
\leqslant k-1$):
\begin{align}\label{Z_FV_all}
& \left[ (\hat{Z}^{FV})^k_{l} (Q^2/\mu^2) \right]^{(1)} = C_F  \,
\frac{\alpha_s}{8\pi} \ln\left( \frac{Q^2}{\mu^2}\right)
\nonumber \\
& \times \left\{ \left( \frac{1}{k} - \frac{2}{k+1} + \frac{2}{k+2}
\right) (k - l) \right. - \frac{1}{k+1} (k - l - 1)
\nonumber \\
& - (-1)^l  \left[ \left( \frac{1}{k} - \frac{2}{k+1} +
\frac{2}{k+2} \right) {k\choose{l+1}} - \left. \frac{1}{k+1}
{k-1\choose{l+1}} \right] \right\} \, .
\end{align}
In particular, it follows from \eqref{Z_FV_all}:
\begin{align}\label{Z_FV_main}
\left[ (\hat{Z}^{FV})^k_{k-1}(Q^2/\mu^2)\right]^{(1)} &= C_F \,
\frac{\alpha_s}{8\pi} \ln\left( \frac{Q^2}{\mu^2}\right) \, [ 1 +
(-1)^k ]
\nonumber \\
& \times \left( \frac{1}{k} - \frac{2}{k+1} + \frac{2}{k+2} \right)
\, .
\end{align}
As one can see from \eqref{Z_FV_main}, $[ (\hat{Z}^{FV})^k_{k-1}
]^{(1)} = 0$, for $k=1$. It is a consequence of the fact that the
singlet operator $\overline \Psi (x) \gamma_{\mu} \!\Psi (x)$ do not
mix with the gluon operators.%
\footnote{Obviously, it should takes place in \emph{all} orders in
$\alpha_s$.}

From \eqref{CF_renorm_rest_4}, \eqref{Z_FV_main} the expression for
the ``major'' gluon CF \eqref{gluon_CF_main} follows which was
obtained in the previous section by solving set of equations
\eqref{gluon_CF_equation}. In order to obtain
$[\tilde{C}^V_{m,l}(Q^2/\mu^2)]^{(1)}$ for $1 \leqslant l \leqslant
m-2$, one should calculate the sum in $k$ in
Eq.~\eqref{CF_renorm_rest_4}. It can be done with the help of
formulae \eqref{Bk02}-\eqref{Bk14} from Appendix~B. As a result, we
come to expressions \eqref{CF_full}, \eqref{CF_even}, \eqref{CF_odd}
derived above on the basis of our main formula \eqref{CF_vs_JJO}.

Finally, by using Eqs.~\eqref{Bl02}-\eqref{Bl08} from Apendix~B, one
can obtain:
\begin{align}\label{sum_kl_main}
\sum_{l=1}^{m-1} & \left[ \tilde{C}^F_{m,m}  \right]^{(0)} \! \left[
(\hat{Z}^{FV})^m_{l} (Q^2/\mu^2) \right]^{(1)} \mathrm{B}(k+2,l+2)
\nonumber \\
& =  C_F \, \frac{\alpha_s}{4\pi} \ln\left( \frac{Q^2}{\mu^2}\right)
[1 + (-1)^m]
\nonumber \\
& \times  \frac{1}{m(m+1)(m+2)} \left\{ \left[ \frac{m-2}{k} +
\frac{2(m+2)}{(k+1)(k+2)} \right. \right.
\nonumber \\
& \hspace{3.8cm} + \left. \frac{m(m+1)+2}{k+m+2} -
\frac{m(m+2)}{k+m+1} \right]
\nonumber \\
& - \left[ (m-2) \mathrm{B}(k, m+3) + m(m+2) \mathrm{B}(k+1, m+2)
\right] \bigg\} \, .
\end{align}
Correspondingly, with the help of formulae \eqref{Bk02}-\eqref{Bk14}
and \eqref{Bl02}-\eqref{Bl08}, we are able to find:
\begin{align}\label{sum_kl_rest}
\sum_{l=1}^{m-1} \sum_{k=l+1}^{m-1} & \!\left[ \tilde{C}^F_{m,k}
\right]^{(0)} \left[ (\hat{Z}^{FV})^k_{l}  (Q^2/\mu^2) \right]^{(1)}
\mathrm{B}(k+2,l+2)
\nonumber \\
& = C_F \, \frac{\alpha_s}{4\pi} \ln\left( \frac{Q^2}{\mu^2}\right)
[1 - (-1)^m]
\nonumber \\
& \times  \frac{1}{m(m+1)(m+2)} \left\{ \left[ \frac{m-2}{k} +
\frac{2(m+2)}{(k+1)(k+2)} \right. \right.
\nonumber \\
& \hspace{3.8cm} + \left. \frac{m(m+1)+2}{k+m+2} -
\frac{m(m+2)}{k+m+1} \right]
\nonumber \\
& - \left[ (m-2) \mathrm{B}(k, m+3) + m(m+2) \mathrm{B}(k+1, m+2)
\right] \bigg\} \, .
\end{align}
Thus, we have successfully reproduced equations \eqref{CF_even} and
\eqref{CF_odd}.

It is well known that any Green function with an insertion of
\emph{one} composite operator is multiplicatively
renormalized~\cite{Zimmermann:73}, while Green functions with
insertion of \emph{two (or more)} composite operators need additive
counterterms~\cite{Collins}. Nevertheless, as was shown in
\cite{Nonsinglet_case}, renormalization group equations for the CFs
have no additive terms, provided the corresponding composite
operators have zero vev.%
\footnote{The renormalization properties of the composite operators
with nonzero vev were studied in \cite{Shore:91}.}

It was found that some \emph{gauge-invariant} singlet composite
operators can mix with \emph{gauge-variant} ones under
renormalization \cite{Joglekar:76}. This problem is present for the
simplest of these operators, the energy-momentum tensor
$\theta_{\mu\nu}$, already in the leading order in strong coupling.

Let $O$ and $N$ represent a set of gauge-independent and
non-gauge-indepen\-dent operators, respectively. It was proven that
renormalized and unrenormalized operators of these types are related
by a triangular matrix \cite{Joglekar:76}:
\begin{equation}\label{triangular_matrix}
  \begin{pmatrix}
  O \\ N
  \end{pmatrix}_{\!\!R} =
  \begin{pmatrix}
  \mathbf{Z}_{OO} & \mathbf{Z}_{ON} \\
   0     & \mathbf{Z}_{NN}
  \end{pmatrix}
 \begin{pmatrix}
  O \\ N
  \end{pmatrix}_{\!\!U} \, ,
\end{equation}
where $\mathbf{Z}_{AB}$ are matrices. In other words, gauge-variant
operators do not mix with gauge-invariant operators under the
renormalization. Correspondingly, anomalous dimensions of
gauge-independent operators can be determined from matrix
$\mathbf{Z}_{OO}$ alone. Moreover, in so-called \emph{physical}
(axial) gauge, $n_{\mu} A^{\mu} = 0$, $n_{\mu}$ being a constant
light-like vector \cite{Bassetto}, a renormalization procedure does
not require gauge-variant counterterms for the gauge-invariant
composite operators at all \cite{Bassetto:94}.

It follows from \eqref{triangular_matrix} that in the OPE
renormalized CFs of gauge-independent operators change under
rescaling of renormalization mass, $\mu_0 \rightarrow \mu$, as
follows:
\begin{equation}\label{rescaling_CF}
C_O (\mu^2) =  C_O (\mu_0^2) \, \mathbf{\hat{Z}}_{OO}(\mu_0^2/\mu^2)
\, ,
\end{equation}
where $ \mathbf{\hat{Z}}_{OO}$ is a matrix of finite
renormalization. All physical matrix elements of gauge-variant
operators vanish \cite{Joglekar:76}.%
\footnote{The calculations made in \cite{Hamberg:92} contradict this
result. However, it was shown by explicit calculations
\cite{Collins:94} that the proof in \cite{Hamberg:92} breaks down,
and conclusions of paper \cite{Joglekar:76} remain true.}
Using a complete set of hadronic states $| n \rangle$, we find that
\begin{equation}\label{gauge_dependent matrix_elements}
\langle N O \rangle = \sum_n \langle N | n \rangle \langle n | O
\rangle = 0 \, .
\end{equation}
Thus, a presence of gauge-variant composite operator in the
OPE~\eqref{OPE} have no influence on our method of determining CFs
in terms of vacuum matrix elements. Indeed, multiplying elements of
the sum $\sum_m [C_O^m \, O_m + C_N^m \, N_m]$ by one of the
gauge-invariant operators, $O_n$, and putting them between vacuum
states, we exclude a contribution from the gauge-variant operators
to our main equation   \eqref{CF_vs_JJO}.

Taking all said above into account, we did not include
gauge-dependent composite operators in the OPE~\eqref{OPE}.


\section{Conclusions and discussions}

As it was shown in the present paper, the singlet CFs of the OPE of
two currents can be explicitly expressed in terms of the Green
functions of the corresponding composite operators without explicit
use of the elementary (quark and gluon) fields. Our main
equation~\eqref{CF_vs_JJO} is a generalization of an analogous
formula which was previously obtained for the singlet case
\cite{Nonsinglet_case}. It is necessary to stress that our formula
holds in \emph{any} renormalization scheme in contrast with other
prescriptions (see, for instance, \cite{Chetyrkin:82}).

As an illustration of a validity of our scheme, the gluon CFs were
calculated in QCD in the first order of the strong coupling
constant. It is important to note that both ``diagonal'' CFs,
$\tilde{C}^g_{m,m-1}(Q^2/\mu^2)$ \eqref{gluon_CF_main}, and
``nondiagonal'' CFs, $\tilde{C}^g_{m,l}(Q^2/\mu^2)$ ($1 \leq l \leq
m-2$) \eqref{CF_full}-\eqref{CF_odd}, were simultaneously obtained.
The renormalization of these composite operators and their CFs were
also considered.

For further discussion, let us rewrite a set of equations for the
singlet quark and gluon CFs in the following symbolic form:
\begin{align}\label{CFs_equations}
\langle J J \, O_q \rangle &  = C_q \langle O_q \, O_q \rangle + C_g
\langle O_g \, O_q \rangle \, , \nonumber \\
\langle J J \, O_g \rangle &  = C_q \langle O_q \, O_g \rangle + C_g
\langle O_g \, O_g \rangle \, .
\end{align}
These equations must be considered as the set of matrix equations
(for simplicity, summations in $l$ are omitted). It is also assumed
that the procedure described in details in Section~\ref{sec:OPE} is
applied to all matrix elements  in \eqref{CFs_equations}.%
\footnote{See derivation of formula \eqref{CF_vs_JJO}.}

\begin{figure}[ht]
\begin{center}
\resizebox{.6\textwidth}{!}{\includegraphics{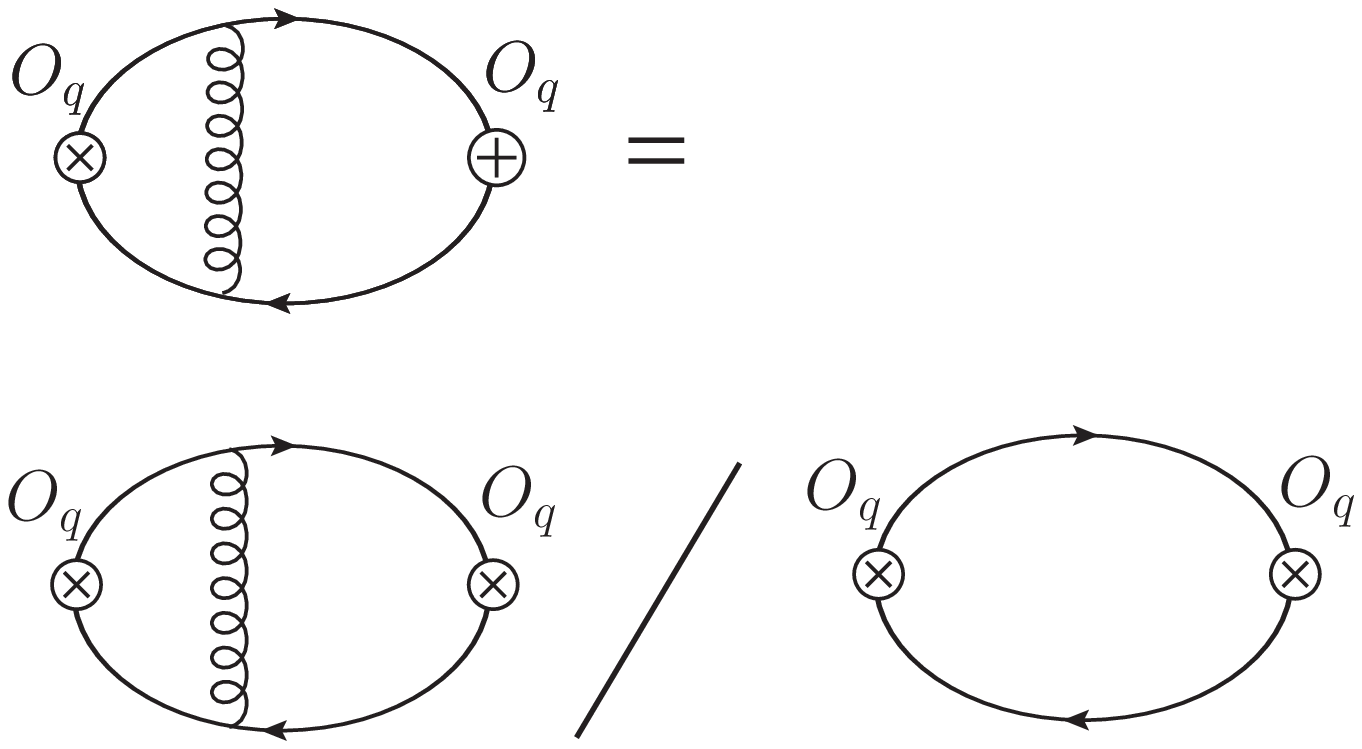}}
\caption{The redefinition of the matrix element of the quark
composite operators in the first order in strong coupling
$\alpha_s$.} \label{fig:reduced_propagator_q-q_1}
\end{center}
\end{figure}

\begin{figure}[ht]
\begin{center}
\resizebox{.6\textwidth}{!}{\includegraphics{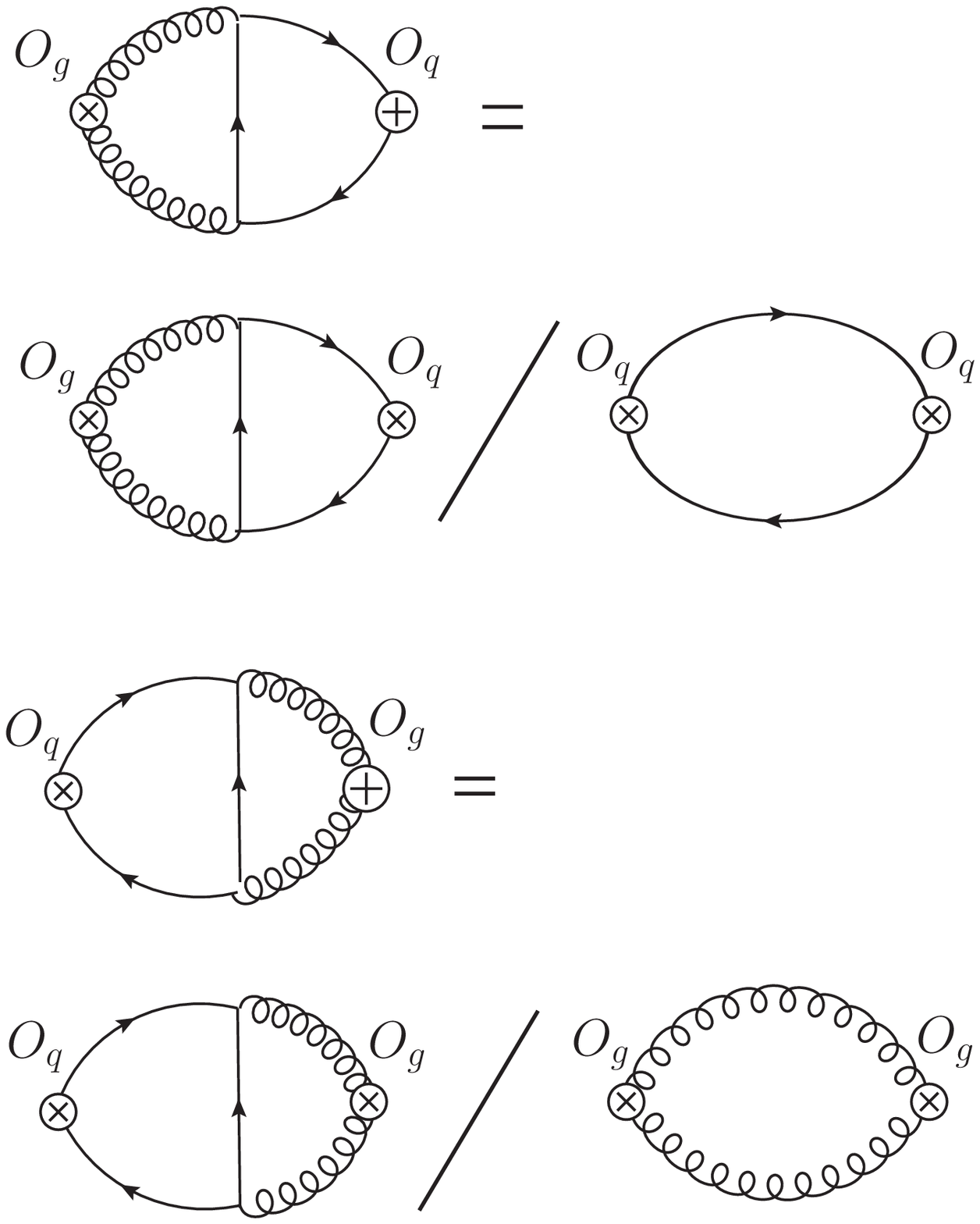}}
\caption{The redefinition of the matrix elements of the quark and
gluon composite operators in the first order in strong coupling
$\alpha_s$.} \label{fig:reduced_propagators_q-g_1}
\end{center}
\end{figure}

\begin{figure}[ht]
\begin{center}
\resizebox{.7\textwidth}{!}{\includegraphics{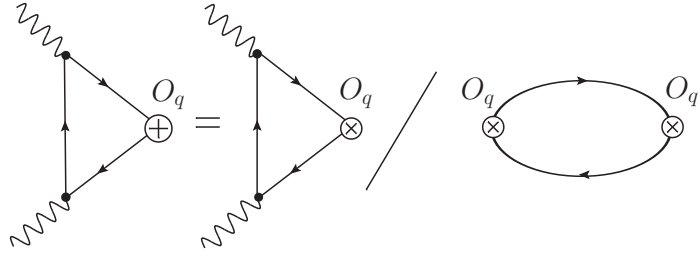}}
\caption{The redefinition of the matrix element $\langle J J \,
O_q\rangle$ in zero order in strong coupling $\alpha_s$.}
\label{fig:reduced_vertex_quark_0}
\end{center}
\end{figure}

\begin{figure}[ht]
\begin{center}
\resizebox{.7\textwidth}{!}{\includegraphics{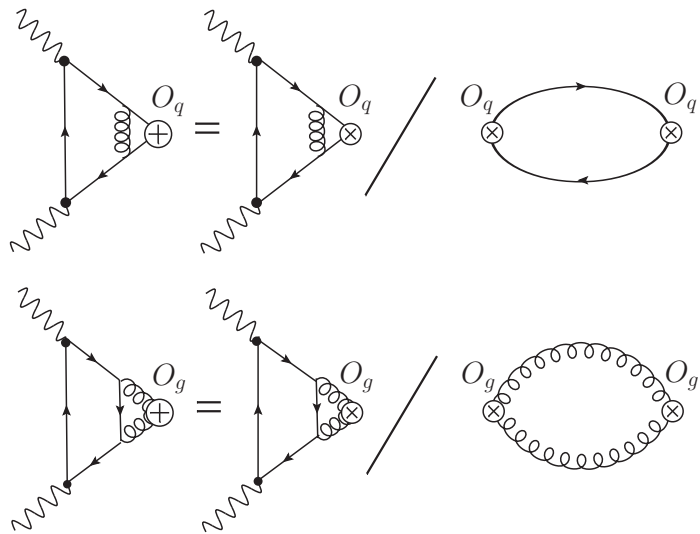}}
\caption{The redefinition of the matrix elements $\langle J J \,
O_q\rangle$ and $\langle J J \, O_g\rangle$ in the first order in
strong coupling $\alpha_s$.} \label{fig:reduced_vertices_q-g_1}
\end{center}
\end{figure}

Let us define ``reduced'' matrix elements of the quark and gluon
composite operators in the $n$-th order of perturbation theory ($n
\geq 1$):
\begin{align}\label{reduced_propagators}
\langle \widehat{O_A \, O_q} \rangle^{(n)} & = \langle O_A \, O_q
\rangle^{(n)} \big[ \langle O_q \, O_q \rangle^{(0)} \, \big]^{-1}
\, ,
\nonumber \\
\langle \widehat{O_A \, O_g} \rangle^{(n)} & = \langle O_A \, O_g
\rangle^{(n)} \, \big[ \langle O_g \, O_g \rangle^{(0)} \big]^{-1}
\, ,
\end{align}
where $A = q, \, g$ (see
Figs.~\ref{fig:reduced_propagator_q-q_1}-\ref{fig:reduced_propagators_q-g_1}).
Analogously, we can define ``reduced'' matrix elements which contain
both composite operators and electromagnetic currents ($n \geq 0$):
\begin{align}\label{reduced_vertices}
\langle \widehat{J J \, O_q} \rangle^{(n)} & = \langle J J O_q
\rangle^{(n)} \, \big[ \langle O_q \, O_q \rangle^{(0)} \big]^{-1}
\, ,
\nonumber \\
\langle \widehat{J J \, O_g} \rangle^{(n)} & = \langle J J \, O_g
\rangle^{(n)} \, \big[ \langle O_g \, O_g \rangle^{(0)} \big]^{-1}
\,
\end{align}
(see
Figs.~\ref{fig:reduced_vertex_quark_0}-\ref{fig:reduced_vertices_q-g_1}).
Note that each of the matrix elements $\langle O_A \, O_B
\rangle^{(n)}$ and $\langle J J O_A \rangle^{(n)}$ has a divergency
related with a divergency of a corresponding Feynman graphs as a
whole.%
\footnote{Note that zero order matrix elements $\langle O_q(p) \,
O_q(-p) \rangle^{(0)}$, $\langle O_g(p) \, O_g(-p) \rangle^{(0)}$,
and $\mathrm{disc}_{(p+q)^2} \langle J(q) J(-(p+q)) O_q(p)
\rangle^{(0)}$ are proportional to $\ln(1/p^2)$ at $p^2 \rightarrow
0$.}
However, these divergences cancel in the \emph{reduced} matrix
elements $\langle \widehat{O_A \, O_B} \rangle^{(n)}$ and $\langle
\widehat{J J O_A} \rangle^{(n)}$.

From \eqref{CFs_equations} we get
\begin{align}\label{CFs_equations_0}
\langle J J \, O_q \rangle^{(0)} &  = C_q^{(0)} \langle O_q \, O_q
\rangle^{(0)} + C_g^{(0)} \langle O_g \, O_q \rangle^{(0)} \, ,
\nonumber \\
\langle J J \, O_g \rangle^{(0)} &  = C_q^{(0)} \langle O_q \, O_g
\rangle^{(0)} + C_g^{(0)} \langle O_g \, O_g \rangle^{(0)} \, .
\end{align}
Since $\langle J J \, O_g \rangle^{(0)} = \langle O_g \, O_q
\rangle^{(0)} = \langle O_q \, O_g \rangle^{(0)} = 0$, while $
\langle O_g \, O_g \rangle^{(0)} \neq 0$, we immediately obtain in
the leading (zero) order in the strong coupling $\alpha_s$:
\begin{align}
C_g^{(0)} & = 0 \, , \label{CF_gluon_0} \\
C_q^{(0)} & = \langle \widehat{J J \, O_q}
\rangle^{(0)} \, . \label{CF_quark_0}
\end{align}

\begin{figure}[ht]
\begin{center}
\resizebox{.7\textwidth}{!}{\includegraphics{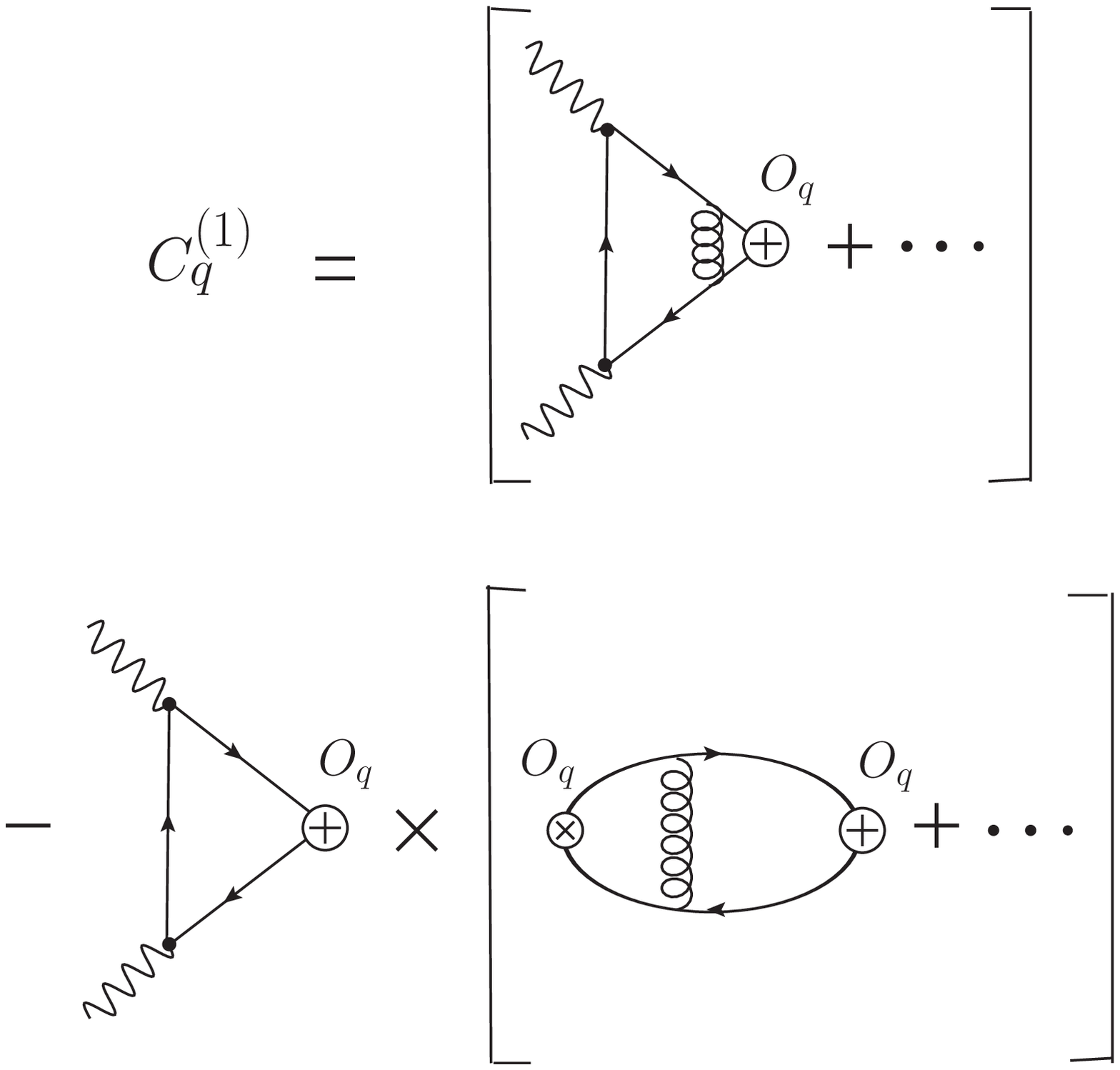}}
\caption{The singlet quark coefficient function of the OPE in the
first order in strong coupling $\alpha_s$.} \label{fig:CF_quark_1}
\end{center}
\end{figure}

\begin{figure}[ht]
\begin{center}
\resizebox{.7\textwidth}{!}{\includegraphics{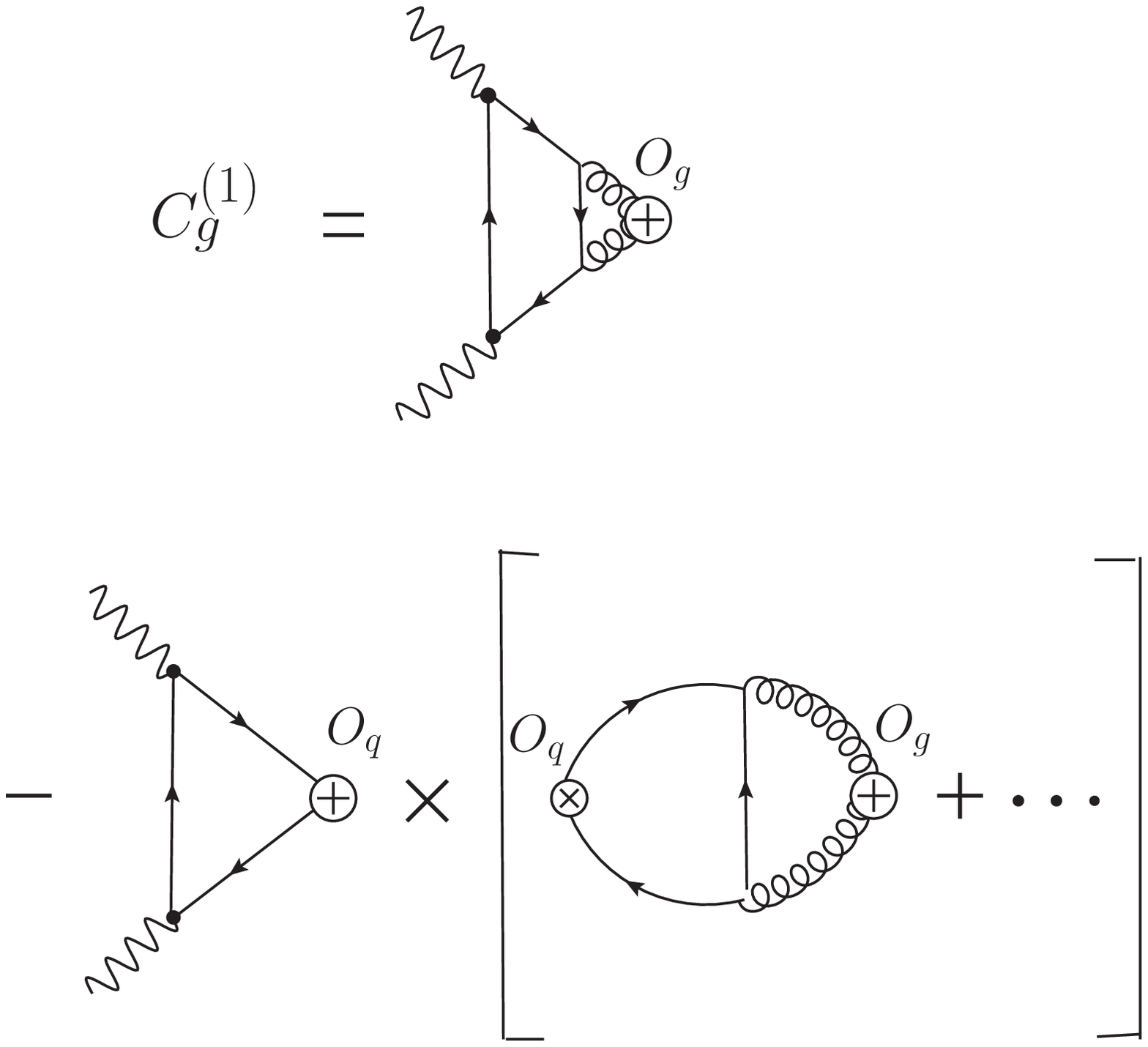}}
\caption{The gluon coefficient function of the OPE in the first
order in strong coupling $\alpha_s$.} \label{fig:CF_gluon_1}
\end{center}
\end{figure}

In the first order in $\alpha_s$, we derive the following
expressions for the singlet quark and gluon CFs:
\begin{align}
C_q^{(1)} = \langle \widehat{J J \, O_q} \rangle^{(1)} & - \langle
\widehat{J J \, O_q} \rangle^{(0)} \langle \widehat{O_q \, O_q}
\rangle^{(1)} \, , \label{CF_quark_1} \\
C_g^{(1)} = \langle \widehat{J J \, O_g} \rangle^{(1)} & - \langle
\widehat{J J \, O_q} \rangle^{(0)} \langle \widehat{O_q \, O_g}
\rangle^{(1)} \, . \label{CF_gluon_1}
\end{align}
The first order contributions to the singlet CFs are presented in
Fig.~\ref{fig:CF_quark_1} and Fig.~\ref{fig:CF_gluon_1},
respectively. In the next order the singlet quark CF looks like
\begin{align}\label{CF_quark_2}
C_q^{(2)} = \langle \widehat{J J \, O_q} \rangle^{(2)} & - \big[
\langle \widehat{J J \, O_q} \rangle^{(1)} - \langle \widehat{J J \,
O_q} \rangle^{(0)} \langle \widehat{O_q \, O_q} \rangle^{(1)} \big]
\langle \widehat{O_q \, O_q} \rangle^{(1)}
\nonumber \\
& - \big[ \langle \widehat{J J \, O_g} \rangle^{(1)} - \langle
\widehat{J J \, O_q} \rangle^{(0)} \langle \widehat{O_q \, O_g}
\rangle^{(1)} \big] \langle \widehat{O_g \, O_q} \rangle^{(1)}
\nonumber \\
& - \langle \widehat{J J \, O_q} \rangle^{(0)} \langle
\widehat{O_q \, O_q} \rangle^{(2)} \, .
\end{align}
Correspondingly, the gluon CF has the form:
\begin{align}\label{CF_gluon_2}
C_g^{(2)} = \langle \widehat{J J \, O_g} \rangle^{(2)} & - \big[
\langle \widehat{J J \, O_q} \rangle^{(1)} - \langle \widehat{J J \,
O_q} \rangle^{(0)} \langle \widehat{O_q \, O_q} \rangle^{(1)} \big]
\langle \widehat{O_q \, O_g} \rangle^{(1)}
\nonumber \\
& - \big[ \langle \widehat{J J \, O_g} \rangle^{(1)} - \langle
\widehat{J J \, O_q} \rangle^{(0)} \langle \widehat{O_q \, O_g}
\rangle^{(1)} \big] \langle \widehat{O_g \, O_g} \rangle^{(1)}
\nonumber \\
& - \langle \widehat{J J \, O_q} \rangle^{(0)} \langle \widehat{O_q
\, O_g} \rangle^{(2)} \, .
\end{align}
In the same way, singlet quark and gluon CFs can be calculated at
any order.

One more advantage of our approach is that it treats uniformly the
CFs of light quarks ($q = u, \, d, \, s$) and CFs of heavy quarks
($Q = c, \, b$). For instance, singlet heavy quark CF, $C_Q$, is
given by the same formulae \eqref{CF_quark_1}, \eqref{CF_quark_2}.
At the same time, the so-called \emph{double counting} problem does
not arise at all.%
\footnote{One possible solution of double counting
problem, which appears in deriving CFs of heavy quarks at the
diagram level, was proposed in \cite{Kisselev:99}.}

The leading parts of the first terms in
Eqs.~\eqref{CF_quark_1}-\eqref{CF_gluon_2}, $\langle \widehat{J J \,
O_A} \rangle^{(n)}$ ($A = q, \, g$) are proportional to
$\ln^n(Q^2/p^2)$, while the leading parts of the matrix elements
$\langle \widehat{O_A \, O_B} \rangle^{(n)}$ ($A, \, B = q, \, g$)
are proportional to $\ln^n(\mu^2/p^2)$. Thus, the role of all terms
in the r.h.s. of Eqs.~\eqref{CF_quark_1}-\eqref{CF_gluon_2}, except
the first one, is \emph{to cancel} $p^2$-\emph{dependence} in the
final expression for the singlet CFs:
\begin{align}
C_A^{(1)} & = N_A^{(1)} \left[ \ln  \!\left( \frac{Q^2}{p^2}\right)
- \ln \!\left( \frac{\mu^2}{p^2}\right) \right] = N_A^{(1)} \ln
\!\left(
\frac{Q^2}{\mu^2}\right) \, , \label{CF_1_log_terms} \\
C_A^{(2)} & = N_A^{(2)} \left[  \ln^2 \!\left(
\frac{Q^2}{p^2}\right) - 2 \ln \!\left( \frac{Q^2}{p^2}\right)  \ln
\!\left( \frac{\mu^2}{p^2} \right)+ \ln^2 \!\left(
\frac{\mu^2}{p^2}\right) \right] \nonumber \\
& = N_A^{(2)} \ln^2 \!\left( \frac{Q^2}{\mu^2}\right) \, ,
\end{align}
where constants $N_A^{(n)}$ are known from explicit calculations.

\begin{figure}[ht]
\begin{center}
\resizebox{.35\textwidth}{!}{\includegraphics{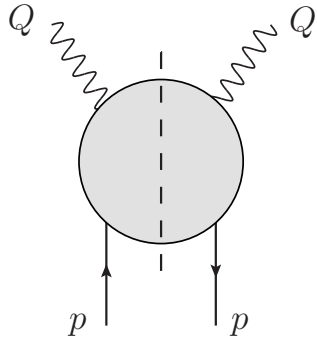}}
\caption{The hadronic part of the amplitude of deep inelastic lepton
scattering off a parton with 4-momentum $p$ ($p^2 < 0$). The dotted
line means that the imaginary part of the amplitude should be
taken.} \label{fig:amplitude}
\end{center}
\end{figure}

An analogy can be drawn between our formulae
\eqref{CF_quark_1}-\eqref{CF_gluon_2} and diagram approach to
calculating CFs.%
\footnote{At the moment, we restrict ourselves by the ``diagonal''
CFs which survive in DIS.}
Namely, the quantity $\langle \widehat{J J \, O_A} \rangle$ should
be associated with the amplitude shown in Fig.~\ref{fig:amplitude},
with $A$ being the type of off-shell parton with 4-momentum $p$. The
$n$-th order contribution to the imaginary part of this amplitude
grows as $\ln^n(Q^2/p^2)$ at $p^2 \rightarrow 0$ (or, equivalently,
at $Q^2 \rightarrow \infty$). The detailed diagram analysis can be
found in Ref.~\cite{Pritchard:80}, where it was shown that the
$\mu^2$-dependence drops out if a gauge-invariant set of QCD
diagrams is taken into account in each order of perturbation theory.

\begin{figure}[ht]
\begin{center}
\resizebox{.24\textwidth}{!}{\includegraphics{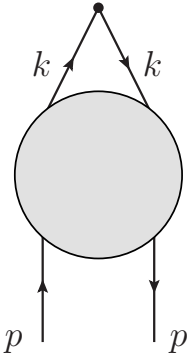}}
\caption{The partonic cut vertex. The solid lines represent quarks
or gluons fields. The ``target'' parton is off-shell, $p^2 < 0$.
Integration in 4-momentum $k$ is made.} \label{fig:cut_vertex}
\end{center}
\end{figure}

The quantity $\langle \widehat{O_A \, O_B} \rangle$ should be
associated with the so-called cut vertex~\cite{Mueller:78} depicted
in Fig.~\ref{fig:cut_vertex}, where upper (lower) lines in
Fig.~\ref{fig:cut_vertex} correspond to partons of type $A(B)$.
Being integrated in 4-momentum $k$ of the upper parton, this diagram
in the $n$-th order has a singularity $\ln^n(\mu^2/p^2)$ at $p^2
\rightarrow 0$.

Our results can be applied to studying \emph{generalized} parton
distributions (GPDs)%
\footnote{Also called \emph{off-forward} or \emph{nonforward} PDs.}
\cite{Muller:94,Radyshkin:96, Ji:97, Hoodbhoy:99} which permanently
attract a great amount of interest.%
\footnote{For the first time, nonforward QCD planar ladder diagrams
were studied in \cite{Bartels:82}.}
They parameterize nonperturbative parton correlation functions in
the nucleons and interpolate between the ordinary parton
distribution functions (PDFs), which can be measured in DIS, and the
elastic form factors. GPDs appear in cross sections of deeply
virtual Compton scattering (DVCS), hard leptoproduction of vector
mesons, as well as in diffractive $Z^0$-production in
$ep$-collision. They were also introduced in the context of the spin
structure of the nucleons~\cite{Ji:97}.

If the OPE \eqref{OPE} is applied to DIS, only ``diagonal''
operators of the type $O^{m}_{F,\,\mu_1 \ldots \mu_m} \! =
O^{m,m}_{F,\,\mu_1 \ldots \mu_m}$ ($O^{m}_{V,\,\mu_1 \ldots \mu_m}
\! = O^{m,m-1}_{V,\,\mu_1 \ldots \mu_m}$) are important, since
forward matrix elements of these operators with $1 \leq l \leq m-1$
($1 \leq l \leq m-2$) are zero. However, for DVCS and other
processes mentioned above, all operators contribute proportionally
to $(p - p')_{\mu_{l+1}} \ldots (p - p')_{\mu_m}$. The invariant
structures of matrix elements of these operators related to the
GPDs. Thus, our scheme of calculating CFs of the ``nondiagonal''
composite operators $ O^{m,l}_{A,\,\mu_1 \ldots \mu_m}$ ($A = F, \,
V$) become quite important.

For the first time, the very notion of nonforward distribution
function was introduced in Ref.~\cite{Kisselev:89}, in which the
statement was made that its Fourier trasformation ``\emph{can be
interpreted as the distribution of partons in momentum fraction} $x$
\emph{and in impact parameter} $b_{\perp}$''. Later on, it was shown
that GPDs in the limit when the momentum transfer is purely
transverse describe the distributions of unpolarized (polarized)
partons in the transverse plane~\cite{Burkardt:03}. Impact parameter
dependent PDFs satisfy positivity constraints which justify their
physical interpretation as probability densities~\cite{Burkardt:03}.

In conclusion, let us stress again that Eq.~\eqref{CF_vs_JJO}, which
defines the singlet CFs in term of the Green functions of the
composite operators, does not apply to perturbation theory at all.
Therefore, our results can be used for calculating CFs of the OPE by
nonperturbative methods.

All Feynman graphs presented in the present paper
(Figs.~\ref{fig:operator_quark_0}-\ref{fig:cut_vertex}) were
prepared with the use of Axodraw package~\cite{Axodraw} and JaxoDraw
graphical user interface~\cite{JaxoDraw}.


\section*{Acknowledgments}

The author would like to thank V.A. Petrov for discussions and
valuable remarks. This work is supported by the grant
RFBR-10-02-00372-a.


\setcounter{equation}{0}
\renewcommand{\theequation}{A.\arabic{equation}}

\section*{Appendix A}
\label{app:B}

In this Appendix we have presented several basic D-dimensional
integrals ($D = 4 - 2\,\varepsilon$) which are needed for our
calculations. Light-cone 4-vector $n_{\mu}$ is defined in the main
text \eqref{LC_vector}. Only \emph{divergent parts} of the integrals
are shown.

\begin{equation}\label{A02}
(\mu^2)^{\varepsilon} \!\! \int \!\! \frac{d^D\!k}{(2\pi)^D}
\frac{(kn)^m}{k^2 (k + p)^2} = \frac{i}{16 \pi^2}
\frac{1}{\varepsilon} \left( \frac{\mu^2}{-p^2}
\right)^{\!\!\varepsilon} \frac{(-1)^m}{m+1} \, (pn)^m \, ,
\end{equation}

\begin{align}\label{A04}
(\mu^2)^{\varepsilon} \!\! \int \!\! \frac{d^D\!k}{(2\pi)^D}
\frac{(kn)^m}{(k+p)^2 (k-l)^2} & = \frac{i}{16 \pi^2}
\frac{1}{\varepsilon} \left[ \frac{\mu^2}{-(l+p)^2}
\right]^{\!\varepsilon} \frac{(-1)^m}{m+1}
\nonumber \\
& \times \sum_{p=0}^{m} (-1)^p (ln)^p \, (pn)^{m-p} \, ,
\end{align}

\begin{align}\label{A06}
(\mu^2)^{\varepsilon} \!\! \int \!\! \frac{d^D\!k}{(2\pi)^D}
\frac{(kn)^m [(k+p)n]^n}{k^2 (k + p)^2} & = \frac{i}{16 \pi^2}
\frac{1}{\varepsilon} \left( \frac{\mu^2}{-p^2}
\right)^{\!\!\varepsilon} (-1)^m  (pn)^{m+n}
\nonumber \\
& \times \mathbf{B}(m+1, \, n+1) \, ,
\end{align}
where $\mathbf{\mathbf{B}}(x, \, y)$ is the beta-function, and $m
\geq 0$. In the next four integrals $m \geq 1$ is assumed:

\begin{align}\label{A12}
(\mu^2)^{\varepsilon} \!\! \int \!\! \frac{d^D\!k}{(2\pi)^D}
\frac{k_{\mu}(kn)^m}{k^2 (k + p)^2} & = \frac{i}{16 \pi^2}
\frac{1}{\varepsilon} \left( \frac{\mu^2}{-p^2}
\right)^{\!\!\varepsilon} \frac{(-1)^{m-1}}{m+2} \, (pn)^{m-1}
\nonumber \\
& \times \left[ - n_{\mu} \, \frac{m}{2(m+1)} \, p^2  + p_{\mu} \,
(pn) \right] \, ,
\end{align}

\begin{align}\label{A14}
(\mu^2)^{\varepsilon} \!\! \int \!\! \frac{d^D\!k}{(2\pi)^D}
\frac{k_{\mu}(kn)^m}{k^2 (k + p)^2 (k-l)^2} & = \frac{i}{32 \pi^2}
\frac{1}{\varepsilon} \left[ \frac{\mu^2}{-(l+p)^2}
\right]^{\!\varepsilon} \frac{(-1)^{m-1}}{m+1} \, n_{\mu}
\nonumber \\
& \times  \sum_{p=0}^{m-1} (-1)^p (ln)^p \, (pn)^{m-p-1} \, ,
\end{align}

\begin{align}\label{A16}
(\mu^2)^{\varepsilon} \!\! \int \!\! \frac{d^D\!k}{(2\pi)^D} &
\frac{k_{\mu}(kn)^m}{(k+p)^2 (k -l)^2} = \frac{i}{16 \pi^2}
\frac{1}{\varepsilon} \left[ \frac{\mu^2}{-(l+p)^2}
\right]^{\!\varepsilon} \frac{(-1)^m}{m+1}
\nonumber \\
& \times \Bigg\{ \frac{1}{2} n_{\mu} (l + p)^2  \Bigg[
\sum_{p=0}^{m-1} (-1)^p (p + 1) (ln)^p \, (pn)^{m-p-1}
\nonumber \\
& \hspace{.5cm} - \frac{1}{m+2} \sum_{p=0}^{m-1} (-1)^p (p + 1)(p +
2) (ln)^p \, (pn)^{m-p-1} \Bigg]
\nonumber \\
& \hspace{.5cm} + (l_{\mu} + p_{\mu}) \, \frac{1}{m+2} \sum_{p=0}^m
(-1)^p (p + 1) (ln)^p \, (pn)^{m-p}
\nonumber \\
& \hspace{.5cm} - p_{\mu} \sum_{p=0}^m (-1)^p (ln)^p \, (pn)^{m-p}
\Bigg\} \, ,
\end{align}

\begin{align}\label{A18}
(\mu^2)^{\varepsilon} \!\! \int \!\! \frac{d^D\!k}{(2\pi)^D} &
\frac{k_{\mu} k_{\nu}(kn)^m}{k^2 (k+p)^2 (k -l)^2} = \frac{i}{32
\pi^2} \frac{1}{\varepsilon} \left[ \frac{\mu^2}{-(l+p)^2}
\right]^{\!\varepsilon} \frac{(-1)^m}{(m+1)(m+2)}
\nonumber \\
& \times \Bigg\{ g_{\mu\nu} \sum_{p=0}^{m} (-1)^p (ln)^p \,
(pn)^{m-p}
\nonumber \\
& \hspace{.5cm} - (l_{\mu} n_{\nu} + n_{\mu} l_{\nu})
\sum_{p=0}^{m-1} (-1)^p (p + 1) (ln)^p \, (pn)^{m-p-1}
\nonumber \\
& \hspace{.5cm} + (p_{\mu} n_{\nu} + n_{\mu} p_{\nu})
\sum_{p=0}^{m-1} (-1)^p (m - p) (ln)^p \, (pn)^{m-p-1}
\nonumber \\
& \hspace{.5cm} - \frac{1}{2} \, n_{\mu} n_{\nu} \Bigg[ k^2
\sum_{p=0}^{m-2} (-1)^p (p + 1) (ln)^p \, (pn)^{m-p-2}
\nonumber \\
& \hspace{2.1cm} + p^2 \sum_{p=0}^{m-2} (-1)^p (m - p - 1) (ln)^p \,
(pn)^{m-p-2}
\nonumber \\
& \hspace{2.1cm} + (l + p)^2 \sum_{p=0}^{m-2} (-1)^p (p + 1) (m - p
- 1)
\nonumber \\
& \hspace{2.1cm} \times (ln)^p \, (pn)^{m-p-2} \Bigg] \Bigg\} \, .
\end{align}

\setcounter{equation}{0}
\renewcommand{\theequation}{B.\arabic{equation}}

\section*{Appendix B}
\label{app:B}

In this Appendix we collected formulae which are needed for
calculating sums presented in the text and getting compact
expressions.%
\footnote{All the formulae collected in this section
were derived by using several table sums with binomial coefficients
\cite{Prudnikov}.} Everywhere below ${n\choose{m}}$ denotes a
binomial coefficient, and $\mathbf{B}(x, y)$ beta-function. For
integer $m, \, n \geqslant 0$, one has $\mathbf{B}(m+1, n+1)
=[(m+n+1) {m+n\choose{n}}]^{-1}$.

Let us first consider summation in index $l$:
\begin{equation}\label{Bkl02}
\sum_{l=1}^{m-1} (-1)^l {m-1\choose{l-1}} \, \mathbf{B} (k+3, l)  =
- (-1)^m \, \mathbf{B} (k+3,m) - \frac{1}{k+m+2}  \, ,
\end{equation}

\begin{align}\label{Bkl04}
\sum_{l=1}^{m-1} (-1)^l & {m-1\choose{l-1}} \frac{1}{l+1} \mathbf{B}
(k+1, l+2)
\nonumber \\
& = - (-1)^m \frac{1}{m+1} \mathbf{B} (k+1,m+2)
\nonumber \\
& - \frac{1}{(k+m+1)(k+m+2)(k+1)} \, ,
\end{align}

\begin{align}\label{Bkl06}
\sum_{l=1}^{m-1} (-1)^l & {m-1\choose{l-1}} \frac{1}{(l+1)(l+2)}
\mathbf{B} (k, l+3)
\nonumber \\
&  = - (-1)^m \frac{1}{(m+1)(m+2)} \mathbf{B} (k,m+3)
\nonumber \\
& - \frac{1}{(k+m+1)(k+m+2)k(k+1)} \, ,
\end{align}

\begin{align}\label{Bkl08}
\sum_{l=1}^{m-1} (-1)^l & {m-1\choose{l-1}} \frac{1}{l(l+1)(l+2)}
\mathbf{B} (k, l+3)
\nonumber \\
& = - (-1)^m \frac{1}{m(m+1)(m+3)} \mathbf{B} (k,m+3)
\nonumber \\
& - \frac{1}{(k+m+2)k(k+1)(k+2)} \, ,
\end{align}

\begin{equation}\label{Bkl10}
\sum_{l=1}^{m-1} (-1)^l {m-1\choose{l-1}} \frac{1}{l+k}  = -
\mathbf{B} (k+1,m) - (-1)^m \frac{1}{m+k} \, .
\end{equation}
In particular, we get from \eqref{Bkl10}:
\begin{equation}\label{Bkl12}
\sum_{l=1}^{m-1} (-1)^l {m-1\choose{l-1}} \frac{1}{l}  = -
\frac{1}{m}[1 +(-1)^m] \, ,
\end{equation}

\begin{equation}\label{Bkl14}
\sum_{l=1}^{m-1} (-1)^l {m-1\choose{l-1}} \frac{1}{l+1}  = -
\frac{1}{m} + \frac{1}{m+1}[1 - (-1)^m] \, ,
\end{equation}

\begin{equation}\label{Bkl16}
\sum_{l=1}^{m-1} (-1)^l {m-1\choose{l-1}} \frac{1}{l+2}  = -
\frac{1}{m+2} (-1)^m - \frac{2}{m(m+1)(m+2)} \, .
\end{equation}

Next four sums in $l$ contain the same beta-function
$\mathbf{B}(k+2, l+2)$:
\begin{equation}\label{Bl02}
\sum_{l=1}^{m-1} (-1)^l  {m\choose{l+1}} \, \mathbf{B} (k+2, l+2)  =
- \frac{m(m-1)}{(k+m+2)(k+2)(k+3)}  \, ,
\end{equation}

\begin{align}\label{Bl04}
\sum_{l=1}^{m-1} (-1)^l {m\choose{l+1}} (m-l-1) \, & \mathbf{B}(k+2,
l+2)
\nonumber \\
&= -\frac{m(m-1)(m-2)}{(k+m+1)(k+2)(k+3)}  \, ,
\end{align}

\begin{equation}\label{Bl06}
\sum_{l=1}^{m-1} \, \mathbf{B} (k+2, l+2) = -  \mathbf{B} (k+1, m+2)
+ \frac{2}{(k+1)(k+2)(k+3)} \, ,
\end{equation}

\begin{align}\label{Bl08}
\sum_{l=1}^{m-1}(l+2) & \mathbf{B} (k+2, l+2) = -  \mathbf{B} (k,
m+3) - (m+2) \mathbf{B} (k+1, m+2)
\nonumber \\
& + (m+1) \mathbf{B} (k+2, m+1) + \frac{6}{(k+1)(k+2)(k+3)} \, .
\end{align}

The following four sums in $k$ are also used during calculations.
The Kronecker symbols $\delta_{n,m}$ garantee that all the sums are
equal to zero for $l=m-1$:
\begin{equation}\label{Bk02}
\sum_{k=l+1}^{m-1} (-1)^l {m\choose{k-1}} \frac{1}{k}
{k\choose{l+1}} = - (-1)^m \frac{1}{m} {m\choose{l+1}} \left[1 -
\delta_{0,m-\,l-1} \right] \, ,
\end{equation}

\begin{align}\label{Bk04}
\sum_{k=l+1}^{m-1} (-1)^l {m\choose{k-1}} \frac{1}{k+1}
{k\choose{l+1}} & = - (-1)^m {m\choose{l+1}} \left[ \frac{1}{m+1} -
\frac{1}{m}\, \delta_{0,m-\,l-1} \right]
\nonumber \\
& + (-1)^l \frac{1}{m(m+1)} \, ,
\end{align}

\begin{align}\label{Bk06}
\sum_{k=l+1}^{m-1} (-1)^l {m\choose{k-1}} \frac{1}{k+2}
{k\choose{l+1}} & = - (-1)^m {m\choose{l+1}} \left[ \frac{1}{m+2} -
\frac{1}{m} \, \delta_{0,m-\,l-1} \right]
\nonumber \\
& + 2(-1)^l \frac{l+2}{m(m+1)(m+2)} \, ,
\end{align}

\begin{align}\label{Bk08}
\sum_{k=l+1}^{m-1} (-1)^l {m\choose{k-1}} \frac{1}{k+1}
{k-1\choose{l+1}} & = - (-1)^m \frac{1}{m+1} {m-1\choose{l+1}}
\nonumber \\
& + (-1)^l  \frac{1}{m} \left[ \frac{l+2}{m+1} - \delta_{0,m-\,l-1}
\right] \, .
\end{align}
Note that the latter sum is equal to zero both for $l=m-1$ and
$l=m-2$.

\begin{equation}\label{Bk10}
\sum_{k=l+1}^{m-1} (-1)^l {m\choose{k-1}} \frac{1}{k}
 = -\frac{1}{m} \left[ (-1)^l {m-1\choose{l}} + (-1)^m \right] \, ,
\end{equation}

\begin{align}\label{Bk12}
\sum_{k=l+1}^{m-1} (-1)^l {m\choose{k-1}} \frac{1}{k+1} & =(-1)^l
\frac{1}{m} \left[ \frac{1}{m+1} {m\choose{l+1}} - {m-1\choose{l}}
\right]
\nonumber \\
&- (-1)^m \frac{1}{m+1} \, ,
\end{align}

\begin{align}\label{Bk14}
\sum_{k=l+1}^{m-1} (-1)^l {m\choose{k-1}} \frac{1}{k+2} & = -(-1)^l
\left[ \frac{1}{m+2} {m+1\choose{l+2}} -
 \frac{2}{m+1} {m\choose{l+2}} \right.
\nonumber \\
&+ \left. \frac{1}{m} {m-1\choose{l+2}} \right] - (-1)^m
\frac{1}{m+2} \, .
\end{align}





\clearpage

\clearpage

\clearpage

\clearpage

\clearpage

\clearpage

\clearpage

\clearpage

\clearpage

\clearpage

\clearpage

\clearpage

\clearpage

\clearpage

\clearpage

\end{document}